%% file: IR_PSZ_cluster.tex
\documentclass[traditabstract, longauth]{aa} 

\usepackage{amsmath}
\usepackage{graphicx}
\usepackage{subcaption}
\usepackage[hyphens]{url}
\usepackage[breaklinks, colorlinks, citecolor=blue]{hyperref}
\usepackage[nonamebreak]{natbib}
\usepackage{longtable,lscape}
\usepackage{txfonts}
\usepackage{epstopdf}
\usepackage{units}
\usepackage{ifthen}
\usepackage{fixltx2e}
\usepackage[stable]{footmisc}
\usepackage{placeins}
\usepackage[table,usenames,dvipsnames]{xcolor}

\input Planck.tex
\providecommand{\sorthelp}[1]{}

\def\reff@jnl#1{{\rm#1\/}}
\def\apj{\reff@jnl{ApJ}}       % Astrophysical Journal dfdfsfsd
\def\apjs{\reff@jnl{ApJS}}     % Astrophysical Journal, Supplement
\def\aaps{\reff@jnl{A\&AS}}    % Astronomy and Astrophysics, Supplement
\def\mnras{\reff@jnl{MNRAS}}   % Monthly Notices of the RAS
\def\prd{\reff@jnl{Phys.\ Rev.\ D}}    % Physical Review D

\newcommand{\beq}{\begin{equation}}
\newcommand{\eeq}{\end{equation}}

\newcommand{\be}{\begin{equation}}
\newcommand{\ee}{\end{equation}}
\newcommand{\bea}{\begin{eq}}
\newcommand{\eea}{\end{equation}}

{\begin{enumerate}\setlength{\itemsep}{0mm}}
{\end{enumerate}}
{\begin{enumerate}\setlength{\itemsep}{0mm}}%
{\end{enumerate}}

\newcommand{\bc}{\begin{center}}
\newcommand{\ec}{\end{center}}
\newcommand{\bi}{\begin{itemize}}
\newcommand{\ei}{\end{itemize}}
\newcommand{\ben}{\begin{enumerate}}
\newcommand{\een}{\end{enumerate}}

\newfont{\gwpfont}{cmssq8 scaled 1000}

\bibpunct{(}{)}{;}{a}{}{,} 

\begin{document}
%%%
%_______________________________________________________________________________
\title{\textit{Planck} intermediate results. XLIII.\\
The spectral energy distribution of dust in clusters of galaxies}

%\input{AuthorList_P05b_SZ_PowerSpectrum_Bispectrum_Py_authors_and_institutes.tex}
\input{PIP_117_Comis_authors_and_institutes.tex}
\date{Received \today \ / Accepted --}

\titlerunning{Dust SED in galaxy clusters}
\authorrunning{Planck Collaboration}

\input{00_abstract}
\maketitle

%______________________________________________________________________________
\section{Introduction}
\label{sec:introduction}
\input{01_introduction}
%______________________________________________________________________________
\section{Data sets}
\label{sec:data}
\input{02_data}
%______________________________________________________________________________
\section{Stacking Analysis}
\label{sec:stacking}
\input{03_stacking}
%______________________________________________________________________________
\section{SED of the cluster IR emission}
\label{sec:results}
\input{04_results}
%_______________________________________________________________________________
\section{Conclusions}
\label{sec:conclusions}
\input{05_conclusions}
%_______________________________________________________________________________
\begin{acknowledgements}
The Planck Collaboration acknowledges the support of: ESA; CNES, and
CNRS/INSU-IN2P3-INP (France); ASI, CNR, and INAF (Italy); NASA and DoE
(USA); STFC and UKSA (UK); CSIC, MINECO, JA, and RES (Spain); Tekes, AoF,
and CSC (Finland); DLR and MPG (Germany); CSA (Canada); DTU Space
(Denmark); SER/SSO (Switzerland); RCN (Norway); SFI (Ireland);
FCT/MCTES (Portugal); ERC and PRACE (EU). A description of the Planck
Collaboration and a list of its members, indicating which technical
or scientific activities they have been involved in, can be found at
{\tt http://www.cosmos.esa.int/web/planck/planck-collaboration}.
This paper makes use of the {\tt HEALPix} software package. We acknowledge the
support of grant ANR-11-BS56-015.
\end{acknowledgements}
%________________________________________________________________________________
\bibliography{Planck_bib,dust_stack}
\bibliographystyle{aat}

\end{document}

%% file: Planck.tex
\def\setsymbol#1#2{\expandafter\def\csname #1\endcsname{#2}}
\def\getsymbol#1{\csname #1\endcsname}

\def\Planck{\textit{Planck}}

\newbox\tablebox    \newdimen\tablewidth
\def\leaderfil{\leaders\hbox to 5pt{\hss.\hss}\hfil}
\def\endPlancktable{\tablewidth=\columnwidth 
    $$\hss\copy\tablebox\hss$$
    \vskip-\lastskip\vskip -2pt}

\def\tablenote#1 #2\par{\begingroup \parindent=0.8em
    \abovedisplayshortskip=0pt\belowdisplayshortskip=0pt
    \noindent
    $$\hss\vbox{\hsize\tablewidth \hangindent=\parindent \hangafter=1 \noindent
    \hbox to \parindent{$^#1$\hss}\strut#2\strut\par}\hss$$
    \endgroup}
\def\doubleline{\vskip 3pt\hrule \vskip 1.5pt \hrule \vskip 5pt}

\def\L2{\ifmmode L_2\else $L_2$\fi}

\def\DeltaT{\ifmmode \Delta T\else $\Delta T$\fi}
\def\deltat{\ifmmode \Delta t\else $\Delta t$\fi}
\def\fknee{\ifmmode f_{\rm knee}\else $f_{\rm knee}$\fi}
\def\Fmax{\ifmmode F_{\rm max}\else $F_{\rm max}$\fi}
\def\solar{\ifmmode{\rm M}_{\mathord\odot}\else${\rm M}_{\mathord\odot}$\fi}
\def\Msolar{\ifmmode{\rm M}_{\mathord\odot}\else${\rm M}_{\mathord\odot}$\fi}
\def\Lsolar{\ifmmode{\rm L}_{\mathord\odot}\else${\rm L}_{\mathord\odot}$\fi}
\def\inv{\ifmmode^{-1}\else$^{-1}$\fi}
\def\mo{\ifmmode^{-1}\else$^{-1}$\fi}
\def\sup#1{\ifmmode ^{\rm #1}\else $^{\rm #1}$\fi}
\def\expo#1{\ifmmode \times 10^{#1}\else $\times 10^{#1}$\fi}
\def\,{\thinspace}
\def\lsim{\mathrel{\raise .4ex\hbox{\rlap{$<$}\lower 1.2ex\hbox{$\sim$}}}}
\def\gsim{\mathrel{\raise .4ex\hbox{\rlap{$>$}\lower 1.2ex\hbox{$\sim$}}}}

\def\simprop{\mathrel{\raise .4ex\hbox{\rlap{$\propto$}\lower 1.2ex\hbox{$\sim$}}}}
\def\deg{\ifmmode^\circ\else$^\circ$\fi}
\def\pdeg{\ifmmode $\setbox0=\hbox{$^{\circ}$}\rlap{\hskip.11\wd0 .}$^{\circ}
          \else \setbox0=\hbox{$^{\circ}$}\rlap{\hskip.11\wd0 .}$^{\circ}$\fi}
\def\arcs{\ifmmode {^{\scriptstyle\prime\prime}}
          \else $^{\scriptstyle\prime\prime}$\fi}
\def\arcm{\ifmmode {^{\scriptstyle\prime}}
          \else $^{\scriptstyle\prime}$\fi}
\newdimen\sa  \newdimen\sb
\def\parcs{\sa=.07em \sb=.03em
     \ifmmode \hbox{\rlap{.}}^{\scriptstyle\prime\kern -\sb\prime}\hbox{\kern -\sa}
     \else \rlap{.}$^{\scriptstyle\prime\kern -\sb\prime}$\kern -\sa\fi}
\def\parcm{\sa=.08em \sb=.03em
     \ifmmode \hbox{\rlap{.}\kern\sa}^{\scriptstyle\prime}\hbox{\kern-\sb}
     \else \rlap{.}\kern\sa$^{\scriptstyle\prime}$\kern-\sb\fi}
\def\ra[#1 #2 #3.#4]{#1\sup{h}#2\sup{m}#3\sup{s}\llap.#4}
\def\dec[#1 #2 #3.#4]{#1\deg#2\arcm#3\arcs\llap.#4}
\def\deco[#1 #2 #3]{#1\deg#2\arcm#3\arcs}
\def\rra[#1 #2]{#1\sup{h}#2\sup{m}}

\def\dots{\relax\ifmmode \ldots\else $\ldots$\fi}
\def\WHzsr{\ifmmode $W\,Hz\mo\,sr\mo$\else W\,Hz\mo\,sr\mo\fi}
\def\mHz{\ifmmode $\,mHz$\else \,mHz\fi}
\def\GHz{\ifmmode $\,GHz$\else \,GHz\fi}
\def\mKs{\ifmmode $\,mK\,s$^{1/2}\else \,mK\,s$^{1/2}$\fi}
\def\muKs{\ifmmode \,\mu$K\,s$^{1/2}\else \,$\mu$K\,s$^{1/2}$\fi}
\def\muKRJs{\ifmmode \,\mu$K$_{\rm RJ}$\,s$^{1/2}\else \,$\mu$K$_{\rm RJ}$\,s$^{1/2}$\fi}
\def\muKHz{\ifmmode \,\mu$K\,Hz$^{-1/2}\else \,$\mu$K\,Hz$^{-1/2}$\fi}
\def\MJysr{\ifmmode \,$MJy\,sr\mo$\else \,MJy\,sr\mo\fi}
\def\MJysrmK{\ifmmode \,$MJy\,sr\mo$\,mK$_{\rm CMB}\mo\else \,MJy\,sr\mo\,mK$_{\rm CMB}\mo$\fi}
\def\microns{\ifmmode \,\mu$m$\else \,$\mu$m\fi}

\def\muK{\ifmmode \,\mu$K$\else \,$\mu$\hbox{K}\fi}
\def\microK{\ifmmode \,\mu$K$\else \,$\mu$\hbox{K}\fi}
\def\muW{\ifmmode \,\mu$W$\else \,$\mu$\hbox{W}\fi}
\def\kms{\ifmmode $\,km\,s$^{-1}\else \,km\,s$^{-1}$\fi}
\def\kmsMpc{\ifmmode $\,\kms\,Mpc\mo$\else \,\kms\,Mpc\mo\fi}
\providecommand{\sorthelp}[1]{}

%% file: PIP_117_Comis_authors_and_institutes.tex
%This author list corresponds to \title{Author list for PIP\_117\_Comis}
%Prepared by M. Lopez-Caniego (Marcos.Lopez.Caniego@sciops.esa.int), ESAC/ESA
%This version is from Fri Mar 11 11:00:25 2016 CET
%\subtitle{There are 179 co-authors in this list}
\author{\small
Planck Collaboration: R.~Adam\inst{73}
\and
P.~A.~R.~Ade\inst{87}
\and
N.~Aghanim\inst{58}
\and
M.~Ashdown\inst{67, 7}
\and
J.~Aumont\inst{58}
\and
C.~Baccigalupi\inst{85}
\and
R.~B.~Barreiro\inst{63}
\and
N.~Bartolo\inst{30, 64}
\and
E.~Battaner\inst{97, 98}
\and
K.~Benabed\inst{59, 95}
% ...
}
\institute{\small
APC, AstroParticule et Cosmologie, Universit\'{e} Paris Diderot, CNRS/IN2P3, CEA/lrfu, Observatoire de Paris, Sorbonne Paris Cit\'{e}, 10, rue Alice Domon et L\'{e}onie Duquet, 75205 Paris Cedex 13, France\goodbreak
\and
Aalto University Mets\"{a}hovi Radio Observatory and Dept of Radio Science and Engineering, P.O. Box 13000, FI-00076 AALTO, Finland\goodbreak
\and
Academy of Sciences of Tatarstan, Bauman Str., 20, Kazan, 420111, Republic of Tatarstan, Russia\goodbreak
\and
African Institute for Mathematical Sciences, 6-8 Melrose Road, Muizenberg, Cape Town, South Africa\goodbreak
\and
Agenzia Spaziale Italiana Science Data Center, Via del Politecnico snc, 00133, Roma, Italy\goodbreak
\and
Aix Marseille Universit\'{e}, CNRS, LAM (Laboratoire d'Astrophysique de Marseille) UMR 7326, 13388, Marseille, France\goodbreak
\and
Astrophysics Group, Cavendish Laboratory, University of Cambridge, J J Thomson Avenue, Cambridge CB3 0HE, U.K.\goodbreak
\and
Astrophysics \& Cosmology Research Unit, School of Mathematics, Statistics \& Computer Science, University of KwaZulu-Natal, Westville Campus, Private Bag X54001, Durban 4000, South Africa\goodbreak
\and
Atacama Large Millimeter/submillimeter Array, ALMA Santiago Central Offices, Alonso de Cordova 3107, Vitacura, Casilla 763 0355, Santiago, Chile\goodbreak
\and
CGEE, SCS Qd 9, Lote C, Torre C, 4$^{\circ}$ andar, Ed. Parque Cidade Corporate, CEP 70308-200, Bras\'{i}lia, DF, Brazil\goodbreak
\and
CITA, University of Toronto, 60 St. George St., Toronto, ON M5S 3H8, Canada\goodbreak
\and
CNRS, IRAP, 9 Av. colonel Roche, BP 44346, F-31028 Toulouse cedex 4, France\goodbreak
\and
California Institute of Technology, Pasadena, California, U.S.A.\goodbreak
\and
Centro de Estudios de F\'{i}sica del Cosmos de Arag\'{o}n (CEFCA), Plaza San Juan, 1, planta 2, E-44001, Teruel, Spain\goodbreak
\and
Computational Cosmology Center, Lawrence Berkeley National Laboratory, Berkeley, California, U.S.A.\goodbreak
\and
Consejo Superior de Investigaciones Cient\'{\i}ficas (CSIC), Madrid, Spain\goodbreak
\and
DSM/Irfu/SPP, CEA-Saclay, F-91191 Gif-sur-Yvette Cedex, France\goodbreak
\and
DTU Space, National Space Institute, Technical University of Denmark, Elektrovej 327, DK-2800 Kgs. Lyngby, Denmark\goodbreak
\and
D\'{e}partement de Physique Th\'{e}orique, Universit\'{e} de Gen\`{e}ve, 24, Quai E. Ansermet,1211 Gen\`{e}ve 4, Switzerland\goodbreak
\and
Departamento de Astrof\'{i}sica, Universidad de La Laguna (ULL), E-38206 La Laguna, Tenerife, Spain\goodbreak
\and
Departamento de F\'{\i}sica, Universidad de Oviedo, Avda. Calvo Sotelo s/n, Oviedo, Spain\goodbreak
\and
Department of Astronomy and Geodesy, Kazan Federal University,  Kremlevskaya Str., 18, Kazan, 420008, Russia\goodbreak
\and
Department of Astrophysics/IMAPP, Radboud University Nijmegen, P.O. Box 9010, 6500 GL Nijmegen, The Netherlands\goodbreak
\and
Department of Physics \& Astronomy, University of British Columbia, 6224 Agricultural Road, Vancouver, British Columbia, Canada\goodbreak
\and
Department of Physics and Astronomy, Dana and David Dornsife College of Letter, Arts and Sciences, University of Southern California, Los Angeles, CA 90089, U.S.A.\goodbreak
\and
Department of Physics and Astronomy, University College London, London WC1E 6BT, U.K.\goodbreak
\and
Department of Physics, Gustaf H\"{a}llstr\"{o}min katu 2a, University of Helsinki, Helsinki, Finland\goodbreak
\and
Department of Physics, Princeton University, Princeton, New Jersey, U.S.A.\goodbreak
\and
Department of Physics, University of California, Santa Barbara, California, U.S.A.\goodbreak
\and
Dipartimento di Fisica e Astronomia G. Galilei, Universit\`{a} degli Studi di Padova, via Marzolo 8, 35131 Padova, Italy\goodbreak
\and
Dipartimento di Fisica e Scienze della Terra, Universit\`{a} di Ferrara, Via Saragat 1, 44122 Ferrara, Italy\goodbreak
\and
Dipartimento di Fisica, Universit\`{a} La Sapienza, P. le A. Moro 2, Roma, Italy\goodbreak
\and
Dipartimento di Fisica, Universit\`{a} degli Studi di Milano, Via Celoria, 16, Milano, Italy\goodbreak
\and
Dipartimento di Fisica, Universit\`{a} degli Studi di Trieste, via A. Valerio 2, Trieste, Italy\goodbreak
\and
Dipartimento di Matematica, Universit\`{a} di Roma Tor Vergata, Via della Ricerca Scientifica, 1, Roma, Italy\goodbreak
\and
Discovery Center, Niels Bohr Institute, Blegdamsvej 17, Copenhagen, Denmark\goodbreak
\and
Discovery Center, Niels Bohr Institute, Copenhagen University, Blegdamsvej 17, Copenhagen, Denmark\goodbreak
\and
European Southern Observatory, ESO Vitacura, Alonso de Cordova 3107, Vitacura, Casilla 19001, Santiago, Chile\goodbreak
\and
European Space Agency, ESAC, Planck Science Office, Camino bajo del Castillo, s/n, Urbanizaci\'{o}n Villafranca del Castillo, Villanueva de la Ca\~{n}ada, Madrid, Spain\goodbreak
\and
European Space Agency, ESTEC, Keplerlaan 1, 2201 AZ Noordwijk, The Netherlands\goodbreak
\and
Gran Sasso Science Institute, INFN, viale F. Crispi 7, 67100 L'Aquila, Italy\goodbreak
\and
HGSFP and University of Heidelberg, Theoretical Physics Department, Philosophenweg 16, 69120, Heidelberg, Germany\goodbreak
\and
Helsinki Institute of Physics, Gustaf H\"{a}llstr\"{o}min katu 2, University of Helsinki, Helsinki, Finland\goodbreak
\and
INAF - Osservatorio Astronomico di Padova, Vicolo dell'Osservatorio 5, Padova, Italy\goodbreak
\and
INAF - Osservatorio Astronomico di Roma, via di Frascati 33, Monte Porzio Catone, Italy\goodbreak
\and
INAF - Osservatorio Astronomico di Trieste, Via G.B. Tiepolo 11, Trieste, Italy\goodbreak
\and
INAF/IASF Bologna, Via Gobetti 101, Bologna, Italy\goodbreak
\and
INAF/IASF Milano, Via E. Bassini 15, Milano, Italy\goodbreak
\and
INFN, Sezione di Bologna, viale Berti Pichat 6/2, 40127 Bologna, Italy\goodbreak
\and
INFN, Sezione di Ferrara, Via Saragat 1, 44122 Ferrara, Italy\goodbreak
\and
INFN, Sezione di Roma 1, Universit\`{a} di Roma Sapienza, Piazzale Aldo Moro 2, 00185, Roma, Italy\goodbreak
\and
INFN, Sezione di Roma 2, Universit\`{a} di Roma Tor Vergata, Via della Ricerca Scientifica, 1, Roma, Italy\goodbreak
\and
INFN/National Institute for Nuclear Physics, Via Valerio 2, I-34127 Trieste, Italy\goodbreak
\and
IPAG: Institut de Plan\'{e}tologie et d'Astrophysique de Grenoble, Universit\'{e} Grenoble Alpes, IPAG, F-38000 Grenoble, France, CNRS, IPAG, F-38000 Grenoble, France\goodbreak
\and
Imperial College London, Astrophysics group, Blackett Laboratory, Prince Consort Road, London, SW7 2AZ, U.K.\goodbreak
\and
Infrared Processing and Analysis Center, California Institute of Technology, Pasadena, CA 91125, U.S.A.\goodbreak
\and
Institut Universitaire de France, 103, bd Saint-Michel, 75005, Paris, France\goodbreak
\and
Institut d'Astrophysique Spatiale, CNRS, Univ. Paris-Sud, Universit\'{e} Paris-Saclay, B\^{a}t. 121, 91405 Orsay cedex, France\goodbreak
\and
Institut d'Astrophysique de Paris, CNRS (UMR7095), 98 bis Boulevard Arago, F-75014, Paris, France\goodbreak
\and
Institute of Astronomy, University of Cambridge, Madingley Road, Cambridge CB3 0HA, U.K.\goodbreak
\and
Institute of Theoretical Astrophysics, University of Oslo, Blindern, Oslo, Norway\goodbreak
\and
Instituto de Astrof\'{\i}sica de Canarias, C/V\'{\i}a L\'{a}ctea s/n, La Laguna, Tenerife, Spain\goodbreak
\and
Instituto de F\'{\i}sica de Cantabria (CSIC-Universidad de Cantabria), Avda. de los Castros s/n, Santander, Spain\goodbreak
\and
Istituto Nazionale di Fisica Nucleare, Sezione di Padova, via Marzolo 8, I-35131 Padova, Italy\goodbreak
\and
Jet Propulsion Laboratory, California Institute of Technology, 4800 Oak Grove Drive, Pasadena, California, U.S.A.\goodbreak
\and
Jodrell Bank Centre for Astrophysics, Alan Turing Building, School of Physics and Astronomy, The University of Manchester, Oxford Road, Manchester, M13 9PL, U.K.\goodbreak
\and
Kavli Institute for Cosmology Cambridge, Madingley Road, Cambridge, CB3 0HA, U.K.\goodbreak
\and
Kazan Federal University, 18 Kremlyovskaya St., Kazan, 420008, Russia\goodbreak
\and
LAL, Universit\'{e} Paris-Sud, CNRS/IN2P3, Orsay, France\goodbreak
\and
LERMA, CNRS, Observatoire de Paris, 61 Avenue de l'Observatoire, Paris, France\goodbreak
\and
Laboratoire AIM, IRFU/Service d'Astrophysique - CEA/DSM - CNRS - Universit\'{e} Paris Diderot, B\^{a}t. 709, CEA-Saclay, F-91191 Gif-sur-Yvette Cedex, France\goodbreak
\and
Laboratoire Traitement et Communication de l'Information, CNRS (UMR 5141) and T\'{e}l\'{e}com ParisTech, 46 rue Barrault F-75634 Paris Cedex 13, France\goodbreak
\and
Laboratoire de Physique Subatomique et Cosmologie, Universit\'{e} Grenoble-Alpes, CNRS/IN2P3, 53, rue des Martyrs, 38026 Grenoble Cedex, France\goodbreak
\and
Laboratoire de Physique Th\'{e}orique, Universit\'{e} Paris-Sud 11 \& CNRS, B\^{a}timent 210, 91405 Orsay, France\goodbreak
\and
Lawrence Berkeley National Laboratory, Berkeley, California, U.S.A.\goodbreak
\and
Lebedev Physical Institute of the Russian Academy of Sciences, Astro Space Centre, 84/32 Profsoyuznaya st., Moscow, GSP-7, 117997, Russia\goodbreak
\and
Max-Planck-Institut f\"{u}r Astrophysik, Karl-Schwarzschild-Str. 1, 85741 Garching, Germany\goodbreak
\and
Moscow Institute of Physics and Technology, Dolgoprudny, Institutsky per., 9, 141700, Russia\goodbreak
\and
National University of Ireland, Department of Experimental Physics, Maynooth, Co. Kildare, Ireland\goodbreak
\and
Nicolaus Copernicus Astronomical Center, Bartycka 18, 00-716 Warsaw, Poland\goodbreak
\and
Niels Bohr Institute, Blegdamsvej 17, Copenhagen, Denmark\goodbreak
\and
Niels Bohr Institute, Copenhagen University, Blegdamsvej 17, Copenhagen, Denmark\goodbreak
\and
Nordita (Nordic Institute for Theoretical Physics), Roslagstullsbacken 23, SE-106 91 Stockholm, Sweden\goodbreak
\and
Optical Science Laboratory, University College London, Gower Street, London, U.K.\goodbreak
\and
SISSA, Astrophysics Sector, via Bonomea 265, 34136, Trieste, Italy\goodbreak
\and
SMARTEST Research Centre, Universit\`{a} degli Studi e-Campus, Via Isimbardi 10, Novedrate (CO), 22060, Italy\goodbreak
\and
School of Physics and Astronomy, Cardiff University, Queens Buildings, The Parade, Cardiff, CF24 3AA, U.K.\goodbreak
\and
Sorbonne Universit\'{e}-UPMC, UMR7095, Institut d'Astrophysique de Paris, 98 bis Boulevard Arago, F-75014, Paris, France\goodbreak
\and
Space Research Institute (IKI), Russian Academy of Sciences, Profsoyuznaya Str, 84/32, Moscow, 117997, Russia\goodbreak
\and
Space Sciences Laboratory, University of California, Berkeley, California, U.S.A.\goodbreak
\and
Special Astrophysical Observatory, Russian Academy of Sciences, Nizhnij Arkhyz, Zelenchukskiy region, Karachai-Cherkessian Republic, 369167, Russia\goodbreak
\and
Sub-Department of Astrophysics, University of Oxford, Keble Road, Oxford OX1 3RH, U.K.\goodbreak
\and
T\"{U}B\.{I}TAK National Observatory, Akdeniz University Campus, 07058, Antalya, Turkey\goodbreak
\and
The Oskar Klein Centre for Cosmoparticle Physics, Department of Physics,Stockholm University, AlbaNova, SE-106 91 Stockholm, Sweden\goodbreak
\and
UPMC Univ Paris 06, UMR7095, 98 bis Boulevard Arago, F-75014, Paris, France\goodbreak
\and
Universit\'{e} de Toulouse, UPS-OMP, IRAP, F-31028 Toulouse cedex 4, France\goodbreak
\and
University of Granada, Departamento de F\'{\i}sica Te\'{o}rica y del Cosmos, Facultad de Ciencias, Granada, Spain\goodbreak
\and
University of Granada, Instituto Carlos I de F\'{\i}sica Te\'{o}rica y Computacional, Granada, Spain\goodbreak
}

%% file: 00_abstract.tex
\abstract {Although infrared (IR) overall dust emission from clusters of galaxies has
been statistically detected using data from the Infrared Astronomical Satellite
(IRAS), it has not been possible to sample the spectral energy distribution
(SED) of this emission over its peak, and thus to break the degeneracy between
dust temperature and mass.  By complementing the IRAS spectral coverage with
\Planck\ satellite data from 100 to 857\,GHz, we provide new constraints on
the IR spectrum of thermal dust emission in clusters of galaxies. 
We achieve this by using a stacking approach for a sample of several hundred
objects from the \Planck\ cluster sample; this procedure averages out
fluctuations from the IR sky, allowing us to reach a significant detection of
the faint cluster contribution.  We also use the large frequency range probed
by \Planck, together with component-separation techniques, to remove the
contamination from both cosmic microwave background anisotropies and the
thermal Sunyaev-Zeldovich effect (tSZ) signal, which dominate at
$\nu\leq353\,$GHz.  By excluding dominant spurious signals or systematic
effects, averaged detections are reported at frequencies
$353\,{\rm GHz}\leq\nu\leq5000\,{\rm GHz}$.  We confirm the presence of dust in
clusters of galaxies at low and intermediate redshifts, yielding an SED with a
shape similar to that of the Milky Way.  \Planck's beam does not allow us
to investigate the detailed spatial distribution of this emission (e.g.,
whether it comes from intergalactic dust or simply the dust content of the
cluster galaxies), but the radial distribution of the emission appears to
follow that of the stacked SZ signal, and thus the extent of the clusters.
The recovered SED allows us to constrain the dust mass responsible for the
signal, as well as its temperature.  We additionally explore the evolution of
the IR emission as a function of cluster mass and redshift.}
\keywords{Galaxies: clusters -- Infrared: general -- Submillimetre: general --
dust}

%% file: 01_introduction.tex
Despite being a minor component of the mass budget of galaxies, dust plays a
significant role from an observational point of view, and might also impact
cosmological studies.
Dust grains can absorb and redistribute light, reddening radiation coming from
background sources, as well as affecting counts of high redshift galaxies
\citep[e.g.,][]{Zwicky1957} or quasars \citep[e.g.,][]{Wright1981, Menard2010}.
Clusters of galaxies are special targets for studying the large-scale
distribution and evolution of dust in the Universe. Furthermore, while the link
between dust and star formation is well established
\citep[][]{Spitzer1978, Sanders1996, Somerville2015}, its role and impact on
the inter-galactic medium (IGM) and intracluster medium (ICM), as well as its
global distribution of properties, are still to be understood.

In clusters of galaxies, the bulk of the baryonic mass is a hot (roughly
$10^7$--$10^8$\,K), ionized, diffuse gas, mostly emitting at X-ray wavelengths
\citep[e.g.,][]{Sarazin1986}. However, since the very beginning, X-ray
spectroscopy has shown the presence of heavy elements within the ICM,
presumably due to stripping of interstellar matter from galaxies, dusty winds
from intracluster stars, and AGN interaction with the ICM
\citep[e.g.,][]{Sarazin1988}. By injecting dust, which in galaxies contains the
bulk of metals, these processes are responsible for the metal enrichment of the
IGM, and the denser and hotter ICM. But the ICM is a hostile environment for
dust grains. The strong thermal sputtering that dust grains undergo at cluster
cores implies lifetimes ranging from $10^{6}$ to $10^{9}$\,yr
\citep{DwekArendt1992}, depending on the gas density and grain size. Thus, with
only the most recently injected material surviving, the cluster dust content is
significantly lower than the typical interstellar values. Despite this,
dust can have a non-negligible role in the cooling/heating of the intracluster
gas, regulated by the properties of the surrounding gas and the radiative
environment \citep{Dwek1990, Popescu2000, Montier2004, Weingartner2006},
and could represent an extra source of non-gravitational physics, influencing
the formation and evolution of clusters and their overall properties
\citep[i.e., cluster scaling relations,][]{daSilva2009}.

Heated by collisions with the hot X-ray emitting cluster gas, ICM dust grains
are expected to emit at far-infrared (FIR) wavelengths
\citep{Dwek1990}. Therefore dust in the
IGM/ICM is expected to contribute to the diffuse IR emission from clusters.
In order to study the effects of cluster environment on the evolution of the member galaxies, 
several studies have been conducted on the dust component of galaxies that are in clusters 
(e.g. \citealt{Braglia2011} with BLAST data, \citealt{Coppin2011} with Herschel data) 
and in massive dark matter halo (\citealt{Welikala2016} $\sim 10^{13}$~M$_{\odot}$ at $z \sim$~1). 
However less has been done on extended dust emission in clusters.
\citet{Stickel2002} used the Infrared Space Observatory (ISO) to look for the
extended FIR emission in six Abell clusters; only towards one of them (A1656,
the Coma cluster) did they find a localized excess of the ratio between the
signal at $120\,\mu$m and $180\,\mu$m, interpreted as being due to thermal
emission from intracluster dust distributed in the ICM, with an approximate
mass estimate of $10^7\,{\rm M}_\odot$. Additionally, although in
qualitative agreement with the ISO result, \citet{Kitayama2009} found only
marginal evidence for this central excess in Coma, based on {\it Spitzer\/}
data.  On the other hand, using a different approach and correlating the Sloan
Digital Sky Survey catalogues of clusters and quasars (behind clusters and in
the field), \cite{Chelouche2007} measured a reddening, typical of dust,
towards galaxy clusters at $z\,{\simeq}\,0.2$, showing that this extinction is
less near the cluster centres, suggesting that most of the detected dust lies
in the outskirts of the clusters.  

A direct study of the IR-emitting dust
in clusters is difficult, because the fluctuations of the IR sky are of
larger amplitude than the flux expected from a single cluster.  However, by
averaging many small patches centred on known cluster positions, a stacking
approach can be used to increase the signal-to-noise ratio, while averaging
down the fluctuations of the IR sky.  This statistical approach was applied
for the first time by \citet{KR1990} considering 71 clusters of galaxies and
IRAS data.  A similar method was also the basis of the detection of the cluster
IR signal reported by \citet{MontierGiard2005}, who exploited the four-band
observations provided by IRAS for a sample of 11\,507 objects.  Later,
\citet{GiardMontier2008} explored the redshift evolution of the IR luminosity
of clusters compared with the X-ray luminosities of the clusters. 
More recently, this approach was used in \citet{planck2014-a29}, where the
correlation between the SZ effect and the IR emission was studied for a
specific sample of clusters. 

When dealing with arcminute resolution data, like those of \Planck\
\footnote{\Planck\ (\url{http://www.esa.int/Planck}) is a project of the
European Space Agency (ESA) with instruments provided by two scientific
consortia funded by ESA member states and led by Principal Investigators from
France and Italy, telescope reflectors provided through a collaboration between
ESA and a scientific consortium led and funded by Denmark, and additional
contributions from NASA (USA).}
and IRAS, the main difficulty for the characterization of the IR properties of
clusters of galaxies is to disentangle the contributions to the overall IR
luminosity coming from the cluster galaxies and that coming from the ICM.  The
overall IR flux is expected to be dominated by the dust emission of the
galaxy component, in particular from star-forming galaxies.
\citet{Roncarelli2010} reconstructed the IRAS stacked flux derived by
\citet{MontierGiard2005} by modelling the galaxy population (using the
SDSS-maxBCG catalogue, consisting of approximately 11\,500 objects with
$0.1\,{<}\,z\,{<}\,0.3$), leaving little room for the contribution of
intra-cluster dust.  Consequently, both the amount of mass in the form of dust
and its location in the clusters, are still open issues.  If the dust
temperature is only poorly constrained, we can obtain only limited constraints
on the corresponding dust mass.  In this work, following the method adopted in
\citet{MontierGiard2005} and \citet{GiardMontier2008}, we combine IRAS and
\Planck\ data, stack these at the positions of the \Planck\ cluster sample
\citep{planck2013-p05a} and investigate the extension and nature of the
corresponding IR signal.  Thanks to the complementary spectral coverage of the
two satellites, we are able to sample the IR emission over its peak in
frequency.  With a maximum wavelength of $100\,\mu$m, IRAS can only explore the
warm dust component, while it is the cold dust that represents the bulk of the
overall dust mass.

The paper is organized as follows. Section~\ref{sec:data} presents the data
used for this study.  We then detail in Sect.~\ref{sec:stacking} the stacking
approach, before discussing the results in Sect.~\ref{sec:results}.  We
summarize and conclude in Sect.~\ref{sec:conclusions}.

%% file: 02_data.tex
\subsection{\textit{Planck} data}
\subsubsection{Frequency maps}
The \Planck\ High Frequency Instrument (HFI) enables us to explore the
complementary side of the IR spectrum (100--857\,GHz), compared with previous
studies based on IRAS data (100--12$\,\mu$m). 
This paper is based on the full (29 month) \Planck-HFI mission, corresponding
to about five complete sky surveys \citep{planck2014-a01}. We use maps from the
six HFI frequency channels (convolved to a common resolution of 10$\arcm$),
pixelized using the {\tt HEALPix} scheme \citep{gorski2005} at
$N_{\rm side}=2048$ at full resolution. 

\subsubsection{Contamination maps}\label{cmb_sz}
While we expect dust emission to be the strongest signal at both 545 and
857\,GHz, at $\nu\leq217\,$GHz the intensity maps will be dominated by
cosmic microwave background (CMB) temperature anisotropies. Furthermore, since
we want to examine known cluster positions on the sky, we must also deal with
the signal from the thermal Sunyaev-Zeldovich (tSZ) effect
\citep{SunyaevZeldovich1,SunyaevZeldovich2}. This latter signal is produced by
the inverse Compton interaction of CMB photons with the hot electrons of the
ICM. The tSZ contribution will be the dominant signal at 100\,GHz and 143\,GHz
(where it shows up as a lower CMB temperature), negligible at 217\,GHz
(since this is close to the SZ null), and also significant at 353\,GHz (where
it shows up as a higher CMB temperature).  Separating the tSZ contribution
from thermal dust emission at $\nu\leq353\,$GHz is a difficult task, which
can only be achieved if the spectrum of the sources is sufficiently well
sampled in the frequency domain. This is the case for the \Planck\ satellite,
whose wide frequency range allows us to reconstruct full-sky maps of both the
CMB and tSZ effect \citep[$y$-map,][]{planck2013-p05b,planck2014-a28} using
adapted component-separation techniques.

These maps have been produced using {\tt MILCA} \citep[the Modified Internal
Linear Combination Algorithm,][]{Hurier2013} with 10$\arcm$ resolution. The
method is based on the well known internal linear combination (ILC) approach
that searches for the linear combination of the input maps, which minimizes the
variance of the final reconstructed signal, under the constraint of offering
unit gain to the component of interest, whose spectral behaviour is known.
{\tt MILCA} relies on the different spatial localization and spectral
properties to separate the different astrophysical components. The quality of the MILCA 
y-map reconstruction has been tested in several ways in the past, comparing different 
flux reconstruction methods, on data and simulations \citep[e.g. ][]{planck2012-V, Hurier2013, planck2014-a28}.

The reconstructed maps have been used to remove the tSZ signal and the CMB
anisotropies, which dominate the stacked maps at frequencies lower than
353\,GHz. The tSZ map also has the advantage of probing the extension of
the clusters, providing a means to check that the IR signal belongs to the
cluster. 

Compared to the publicly released CMB map (specifically {\tt SMICA}, available
through the Planck Legacy Archive\footnote{\url{http://pla.esac.esa.int}}),
the reconstruction obtained
with {\tt MILCA}, imposing conditions to preserve the CMB and remove the SZ
effect, allows one to obtain a more robust extraction of the CMB signal at each
cluster position. The {\tt SMICA} CMB map is, however, more reliable at large
angular scales and has been used to test the robustness of the CMB
reconstruction at scales of $1\deg$. The {\tt SMICA} map has allowed us to
validate the {\tt MILCA} map in the region around each cluster position
(0\pdeg5).

\subsubsection{Cluster sample: the Planck Catalogue of SZ Sources}
\label{cl_sample}
In order to stack at known cluster positions in rather clean sky regions
(e.g., with low Galactic dust contamination), we consider the clusters listed
in the Planck Catalogue of SZ sources \citep{planck2013-p05a}.
For such high significance SZ-detected clusters (${\rm SNR}\,{>}\,4.5$)
belonging to \Planck\ SZ catalogues, the robustness of the tSZ flux
reconstruction has been already investigated and tested
\citep{planck2013-p05b,planck2014-a28}. We expect to be able to subtract the
tSZ signal with high accuracy ($\ll\,10\,\%$), which is necessary in order to
reconstruct the spectral energy distribution (SED) of the IR emission from
clusters at $\nu\leq353\,$GHz. 

The SZ catalogue constructed from the total intensity data taken during the
first 15.5 months of \Planck\ observations
\citep[][PSZ1 hereafter,]{planck2013-p05a} contains 1227 clusters and cluster
candidates. Unlike for the Second Catalogue of Planck SZ sources
\cite[PSZ2,]{planck2014-a36}, the validation process and extensive follow-up
observations for PSZ1 have already been completed, as detailed in
\citet{planck2015-XXXVI}, \citet{planck2014-XXVI}, and
\citet{planck2013-p05a-addendum}. We thus limit our analysis to PSZ1, in which
redshifts and associated mass estimates (derived from the $Y_z$ mass
proxy, as detailed in section 7.2.2 of \citealt{planck2013-p05a}),
are also available for 913 objects.

\subsection{IRAS data}
To sample the thermal dust SED across its peak and to constrain its shape,
we complement the \Planck\ spectral coverage with the 100 and 60-$\mu$m IRAS
maps. We explicitly use the Improved Reprocessing of the IRAS Survey maps
\citep[IRIS,][]{IRIS} \footnote{\url{http://www.ias.fr/IRIS/IrisDownload.html}},
for which artefacts such as zero level, calibration, striping, and residual
zodiacal light have been corrected. The IRIS maps (like previous generation
IRAS maps) are a mosaic of 430 tangent plane projections, covering the whole
sky with 1\parcm5 pixels.  Here, we have used the corresponding sky maps
provided in {\tt HEALPix} format, with $N_{\rm side}=2048$. For the purposes
of the present work, the IRIS 100 and $60\,\mu$m maps are convolved to a
resolution of 10$\arcm$ in order to match that of the \Planck\ frequency and
tSZ maps.

%% file: 03_stacking.tex
\begin{figure*}
  \centering
    \includegraphics[width=0.33\textwidth]{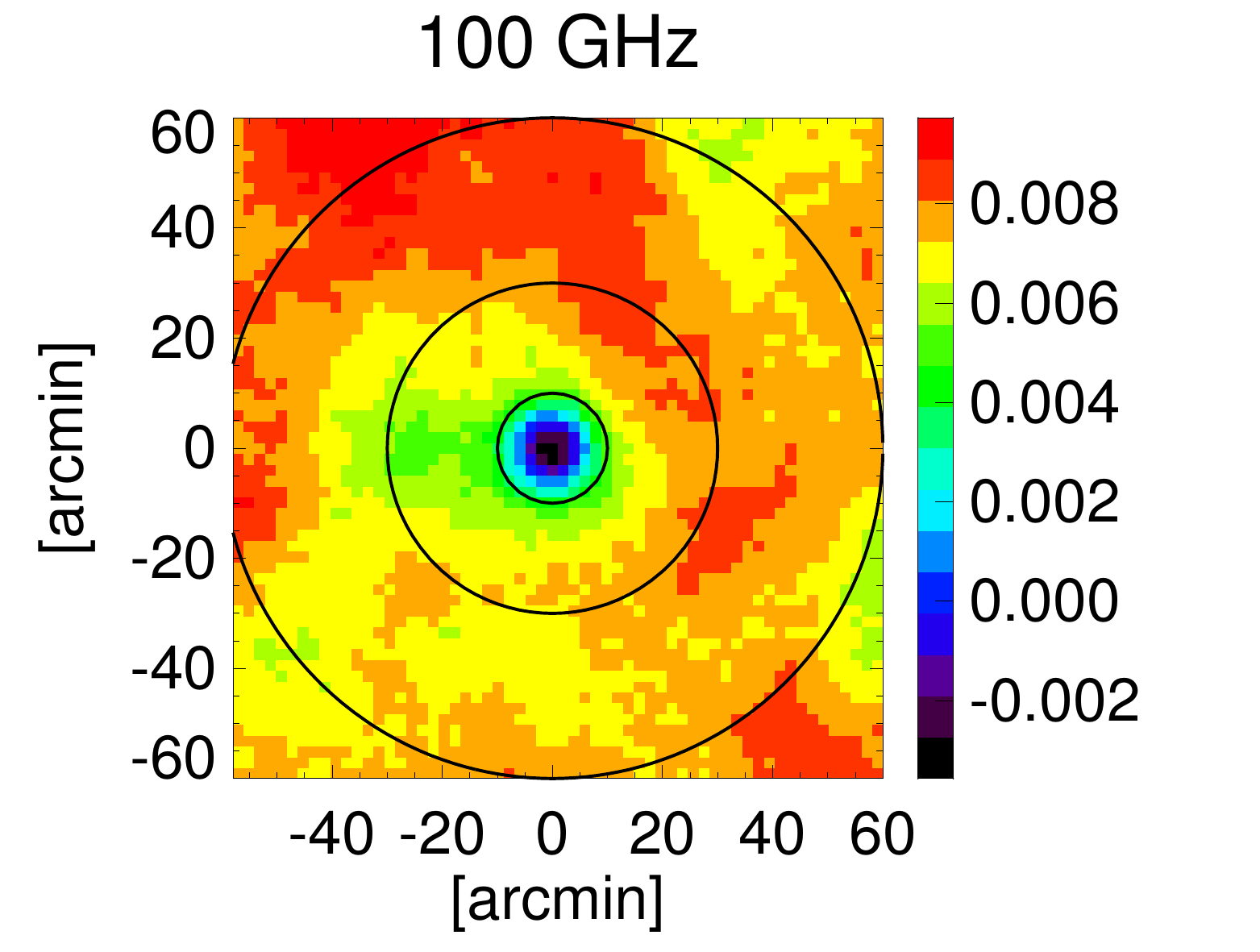}
    \includegraphics[width=0.33\textwidth]{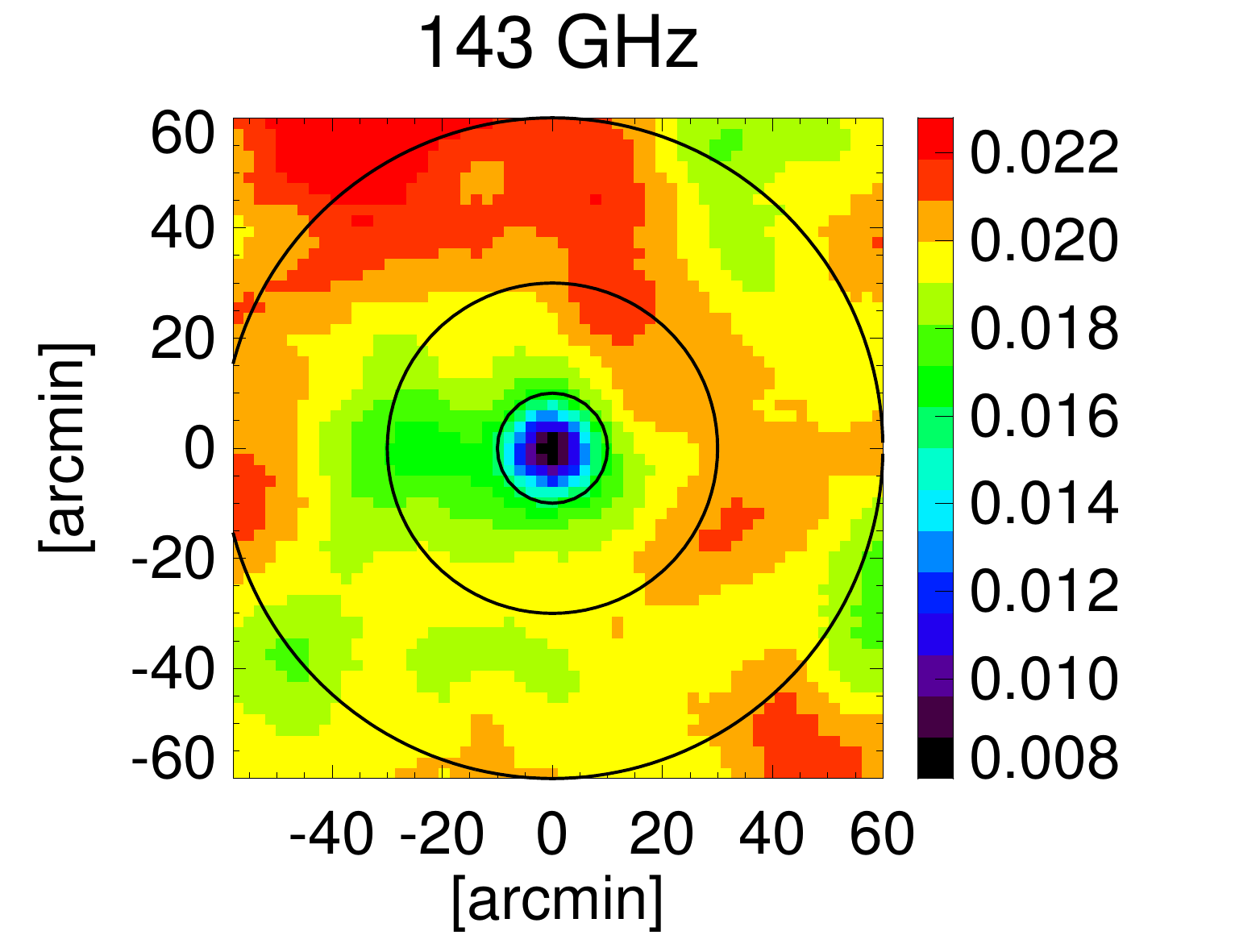}
    \includegraphics[width=0.33\textwidth]{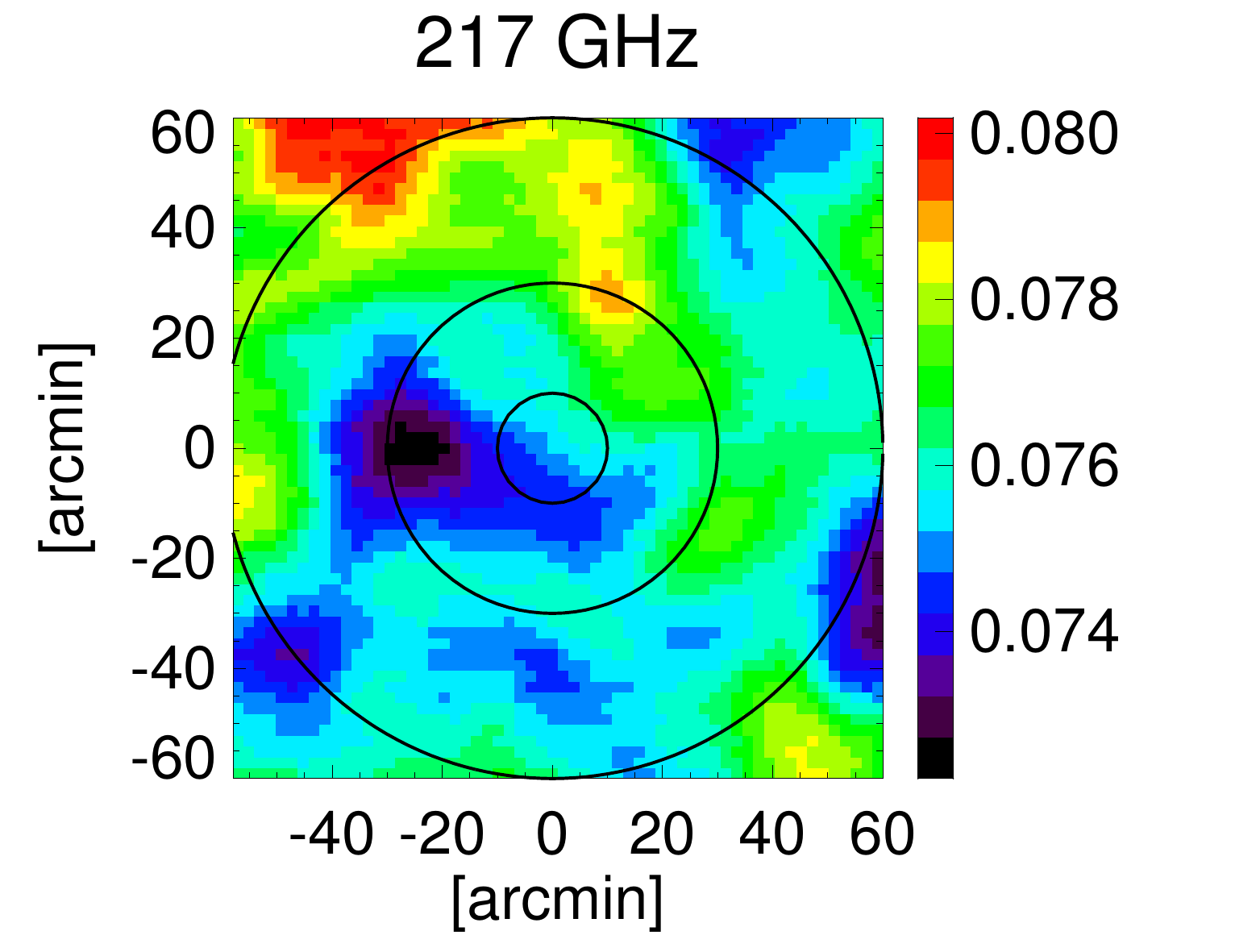}
    \includegraphics[width=0.33\textwidth]{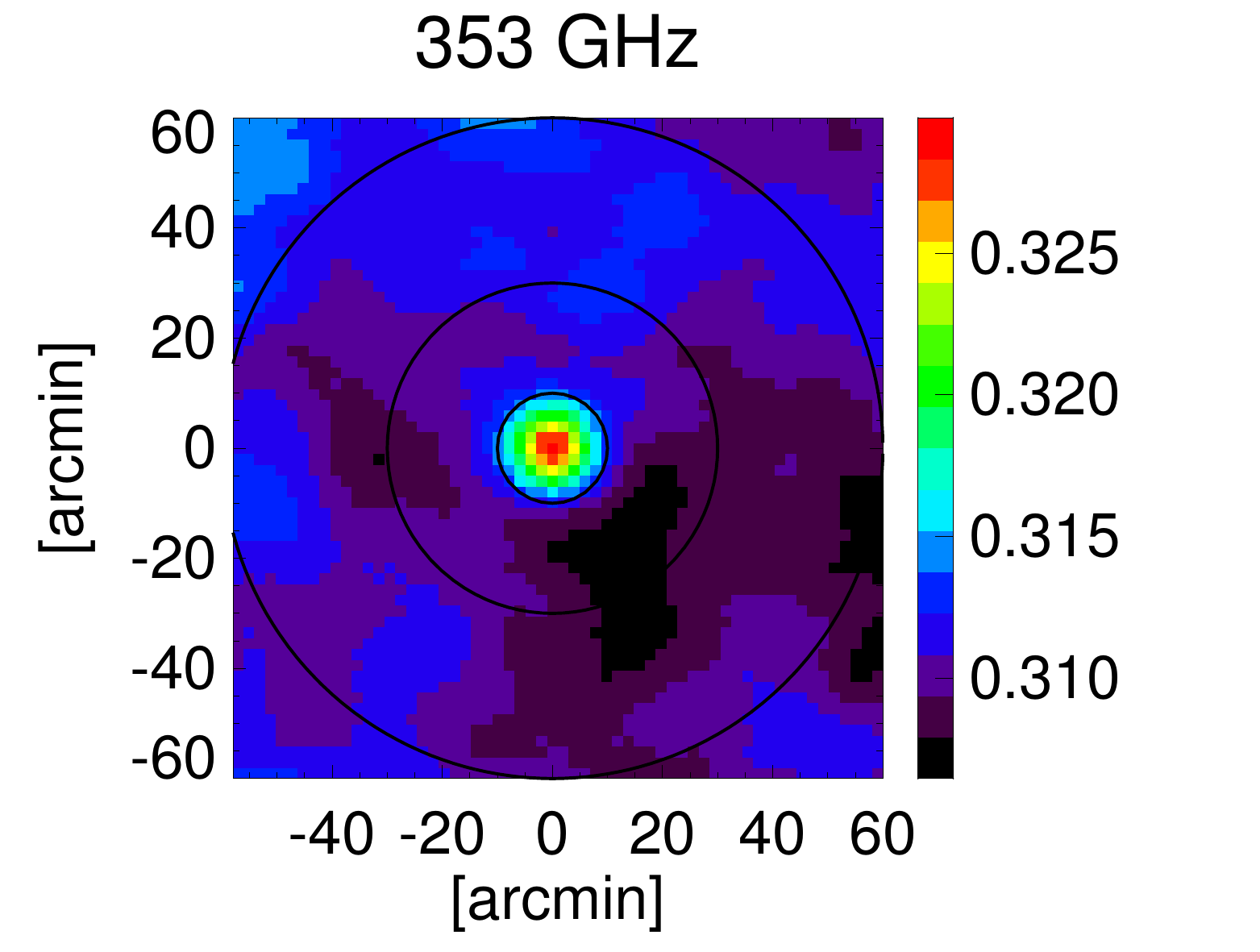}
    \includegraphics[width=0.33\textwidth]{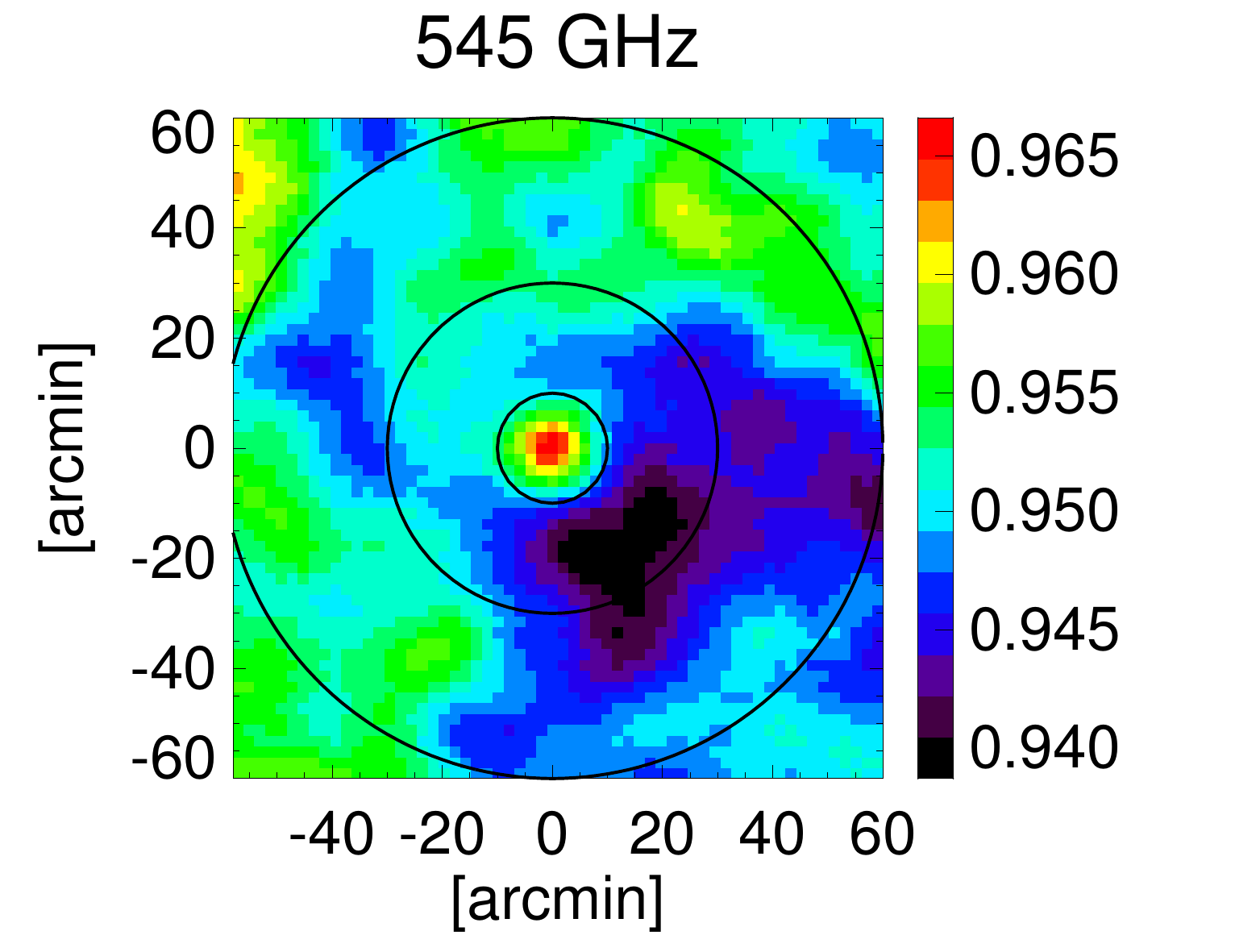}
    \includegraphics[width=0.33\textwidth]{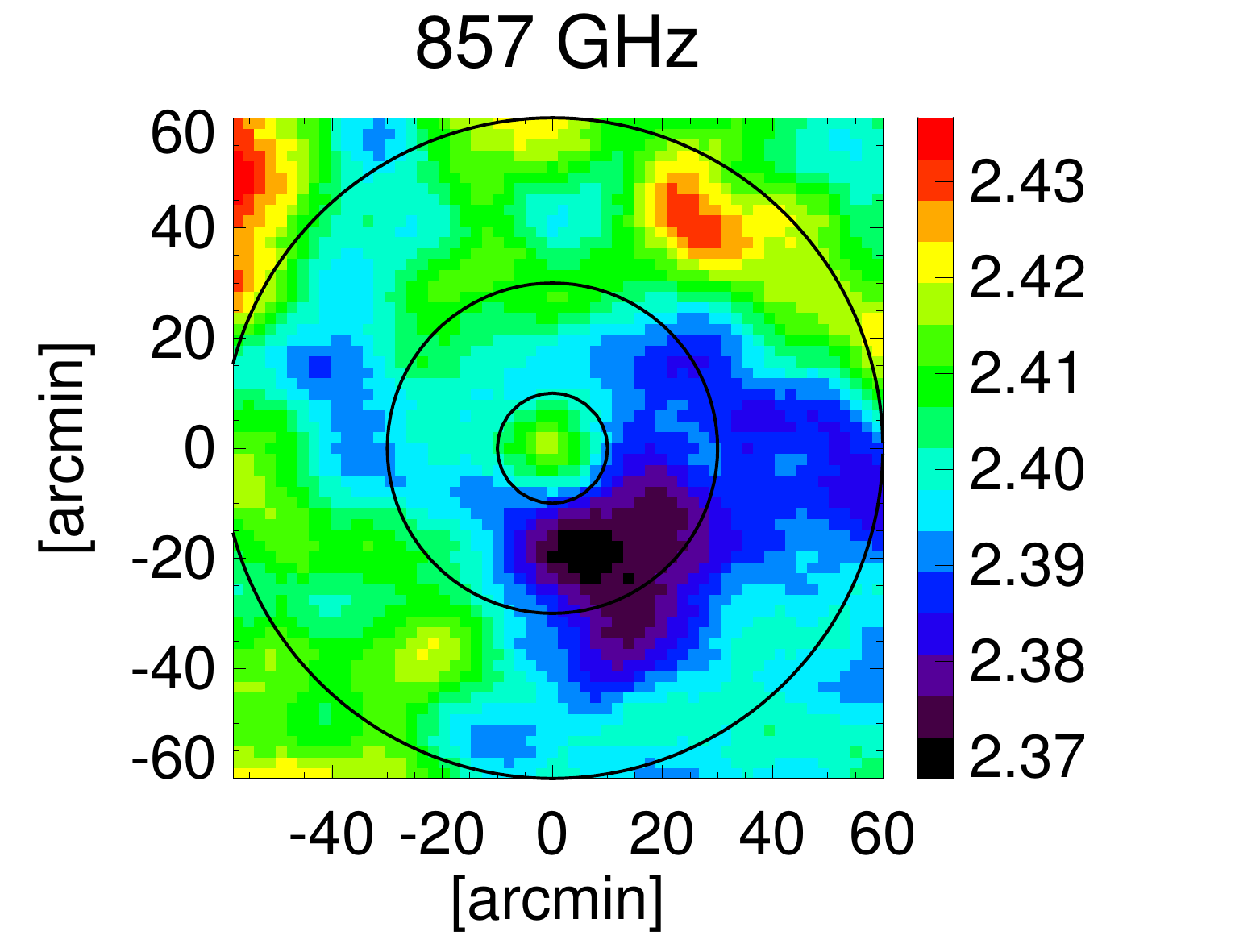}
      \caption{\Planck\ stacked maps at the positions of the final sample of
645 clusters. The maps are $2\deg\times2\deg$ in size and are in units of
${\rm MJy}\,{\rm sr}^{-1}$. Black contours represent circles with a radius
equal to 10$\arcm$, 30$\arcm$, and 60$\arcm$.}
         \label{noclean}
\end{figure*}

Because the sky fluctuations from the cosmic infrared background (CIB)
are stronger than the brightness
expected from a single object, we cannot detect the dust contribution to the
IR emission from individual clusters of galaxies.
However, we can statistically detect the population of clusters by
averaging local maps centred at known cluster positions, thereby
reducing the background fluctuations
\citep[see e.g.,][]{MontierGiard2005, GiardMontier2008}.
Here we take advantage of the ``IAS stacking library''
\citep{IAS_stack2, IAS_stack} in order to co-add cluster-centred regions and
increase the statistical significance of the IR signal at each frequency.
Patches of $2\deg\times2\deg$, centred on the cluster positions
(using 2\arcm\ pixels) have been extracted for the six \Planck-HFI
frequency channels, as well as for the 100 and 60-$\mu$m IRIS maps. 
To ensure a high signal-to-noise ratio and low level of contamination, the low
reliability (``category 3'') PSZ1 cluster candidates (126 objects) have been
excluded from the analysis.
 
We now detail the different steps of the stacking approach that we adopt to
perform our analysis. This includes extraction of the maps at each frequency,
foreground removal, and selection of the final cluster sample.

\subsection{Field selection}\label{field_selection}
\subsubsection{Exclusion of CO contaminated regions}\label{co}
The 100, 217, and 353-GHz \Planck\ channels can be significantly contaminated
by the signal due to the emission of CO rotational transition lines at 115,
230, and 345\,GHz, respectively \citep{planck2013-p03d}. Component separation
methods have been used to reconstruct CO maps from \Planck\ data
\citep{CO2013}. Since the CO emission can be an important foreground for the
purpose of the present work, we choose to use a quite strict CO mask. This
mask is based on the released CO $J\,{=}\,1\,{\rightarrow}\,0$
\Planck\ map \citep{CO2013} and is obtained by applying a
$3\,{\rm K}_{\rm RJ}\,{\rm km}\,{\rm s}^{-1}$ cut on the map (where the
$K_{\rm RJ}$ unit comes from intensity scaled to temperature using the
Rayleigh-Jeans approximation).
We conservatively exclude from the stacking procedure all the fields in which
we find flagged pixels, according to the CO mask, at cluster-centric distances
$\leq1\deg$.  This leads to the exclusion of 55 extra clusters, living us with
1046 remaining cluster positions.

\subsubsection{Point source mask}\label{ps_mask}
Before proceeding to stack the cluster-centred fields, we check that they are
characterized by comparable background contributions. 
As a first step we verify the presence of known point sources, using the masks
provided by the \Planck\ Collaboration for the six frequencies considered here,
as well as the IRAS point source catalogue \citep{1988iras....7.....H}. We
exclude all fields in which point sources are found at cluster-centric
distances $\leq5\arcm$, even if this is the case only for a single channel.
For point sources at larger distances from the nominal cluster position we set
the corresponding pixel values to the mean for the pixels within the
$0\pdeg5\leq r<1\deg$ region of the $2\arcm\times2\deg$ patch, at each
wavelength.  We also check that all the selected cluster fields have a variance
in the $0\pdeg5\leq r<1\deg$ region that is $\leq5$ times that of the whole
sample.  These additional cuts lead to a reduced sample of 645 clusters, with
only two clusters lying at Galactic latitudes lower than 10\deg.
We have also tested the more conservative choice of excluding {\it all\/}
the fields in which known point sources are found at ${\leq}\,10\arcm$ from the
centre.  This substantially reduces the sample size (to 504 clusters), while
giving no significant difference in the recovered signal, as will be discussed
later.
 
\begin{figure*}
  \centering
    \includegraphics[width=0.33\textwidth]{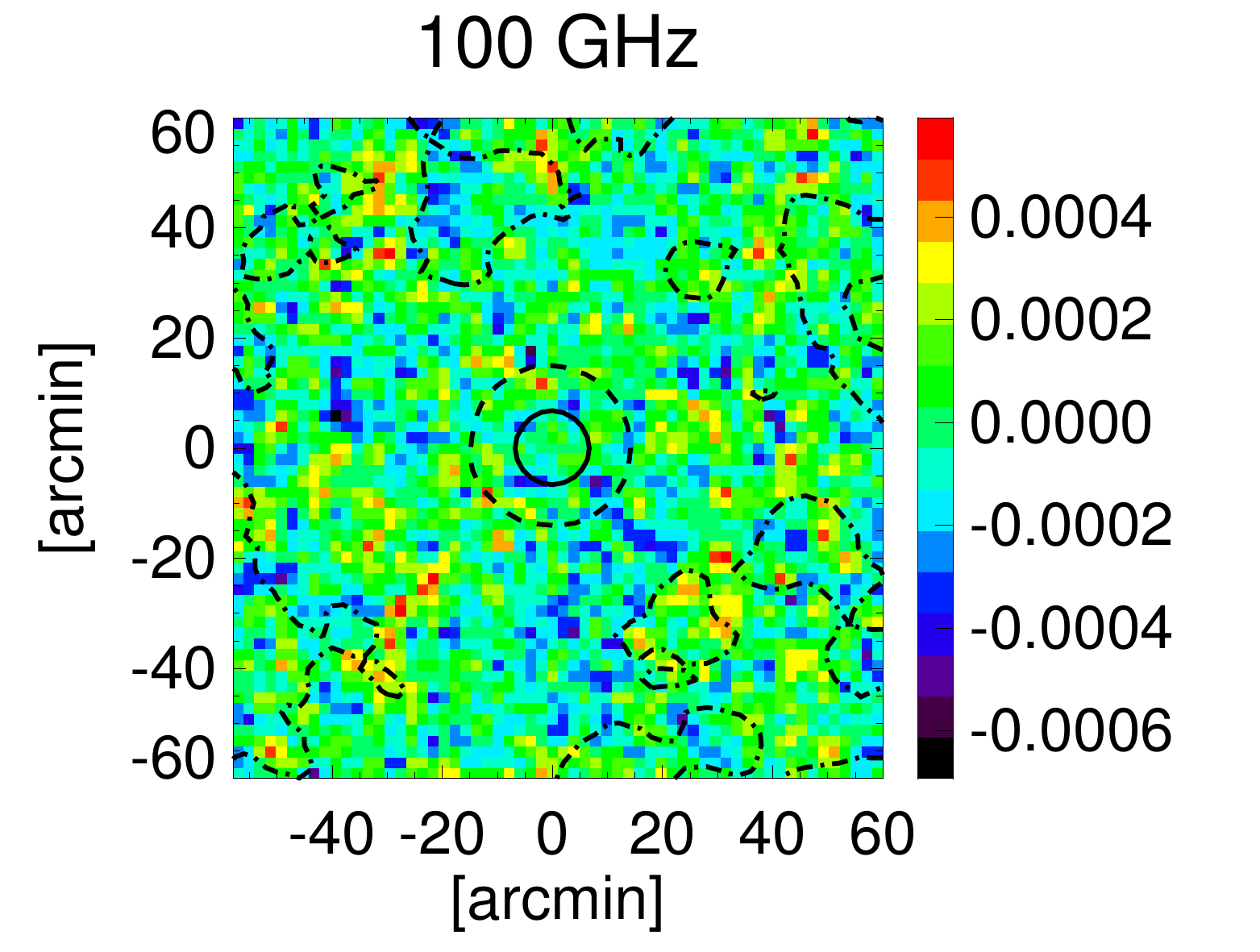}
    \includegraphics[width=0.33\textwidth]{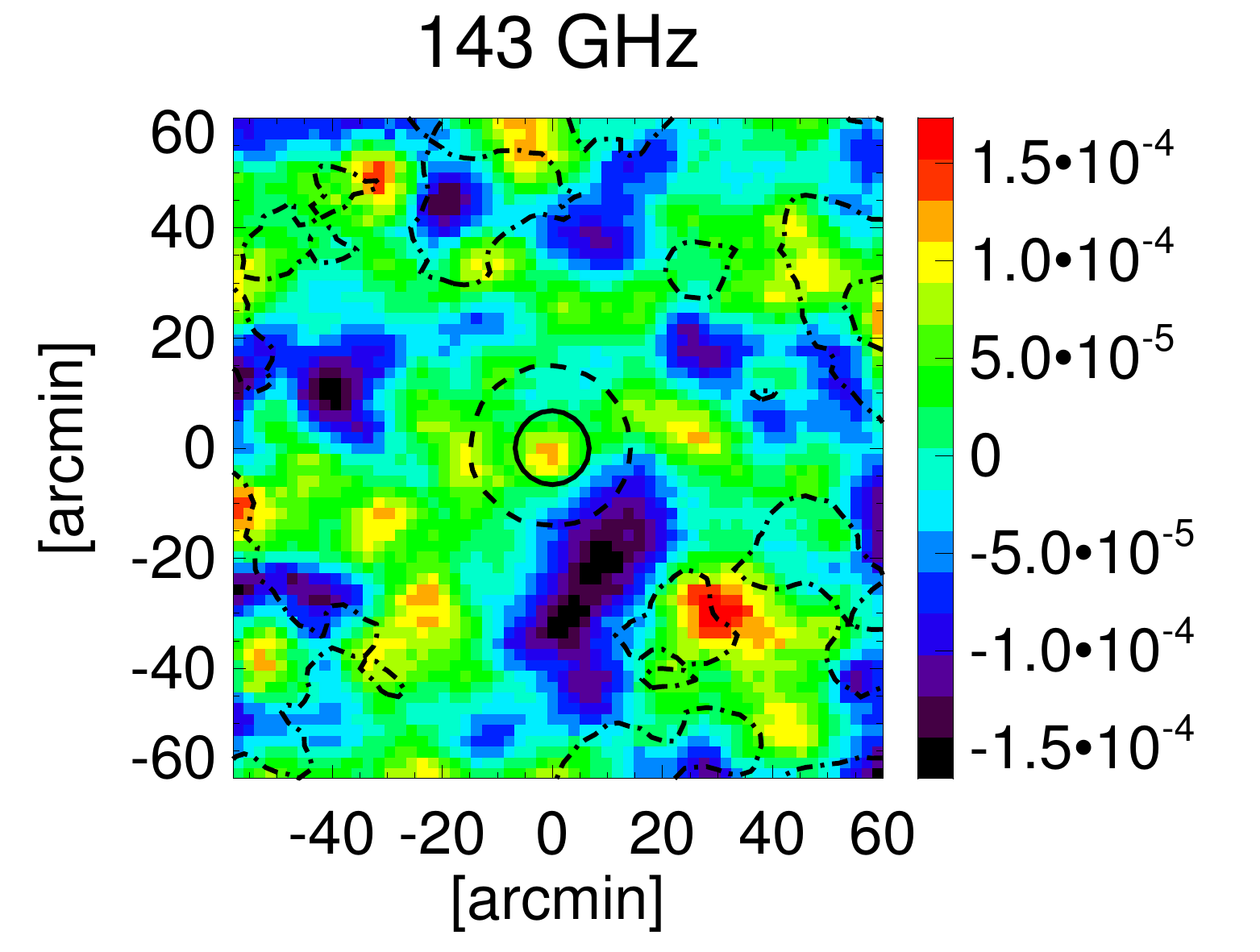}
    \includegraphics[width=0.33\textwidth]{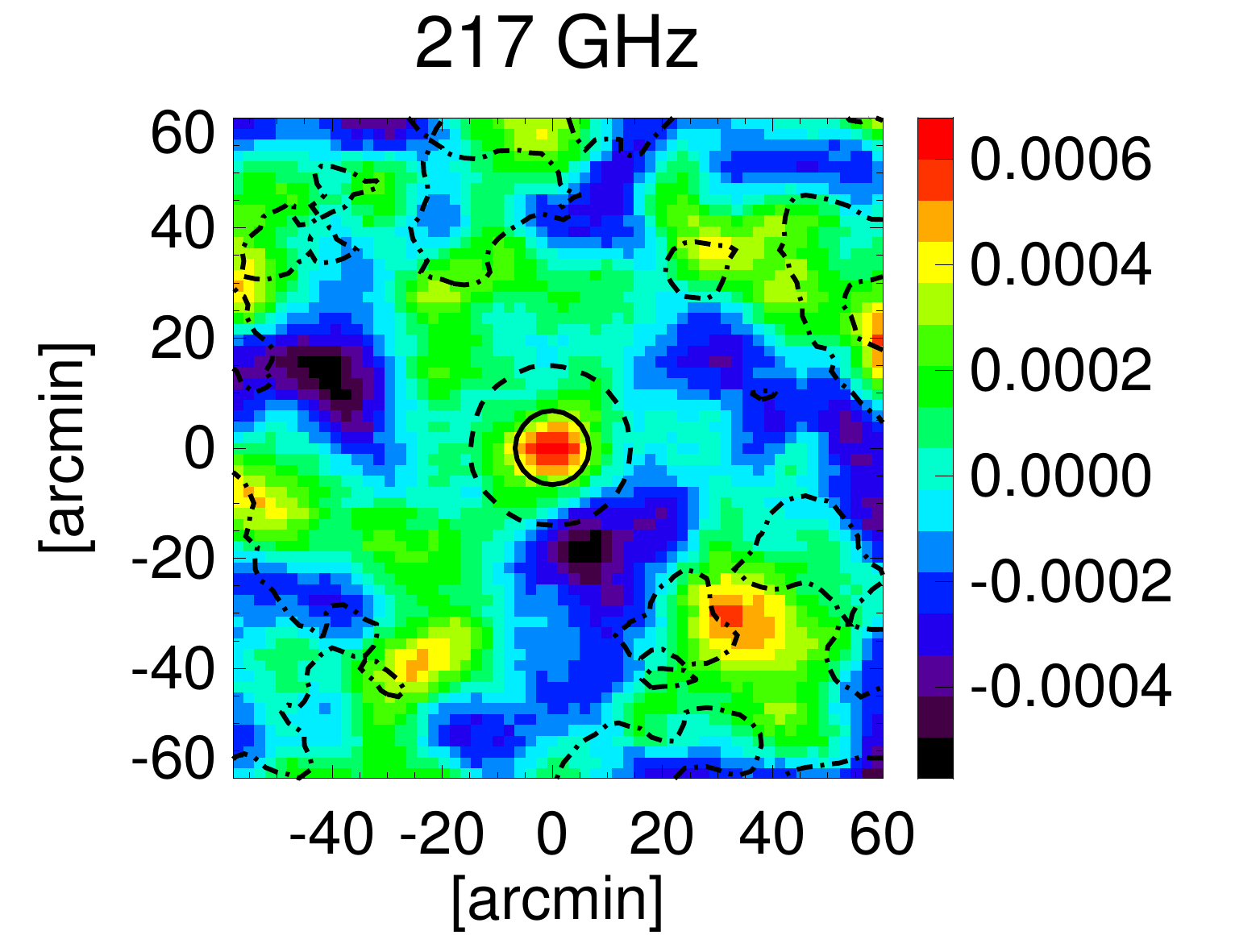}
    \includegraphics[width=0.33\textwidth]{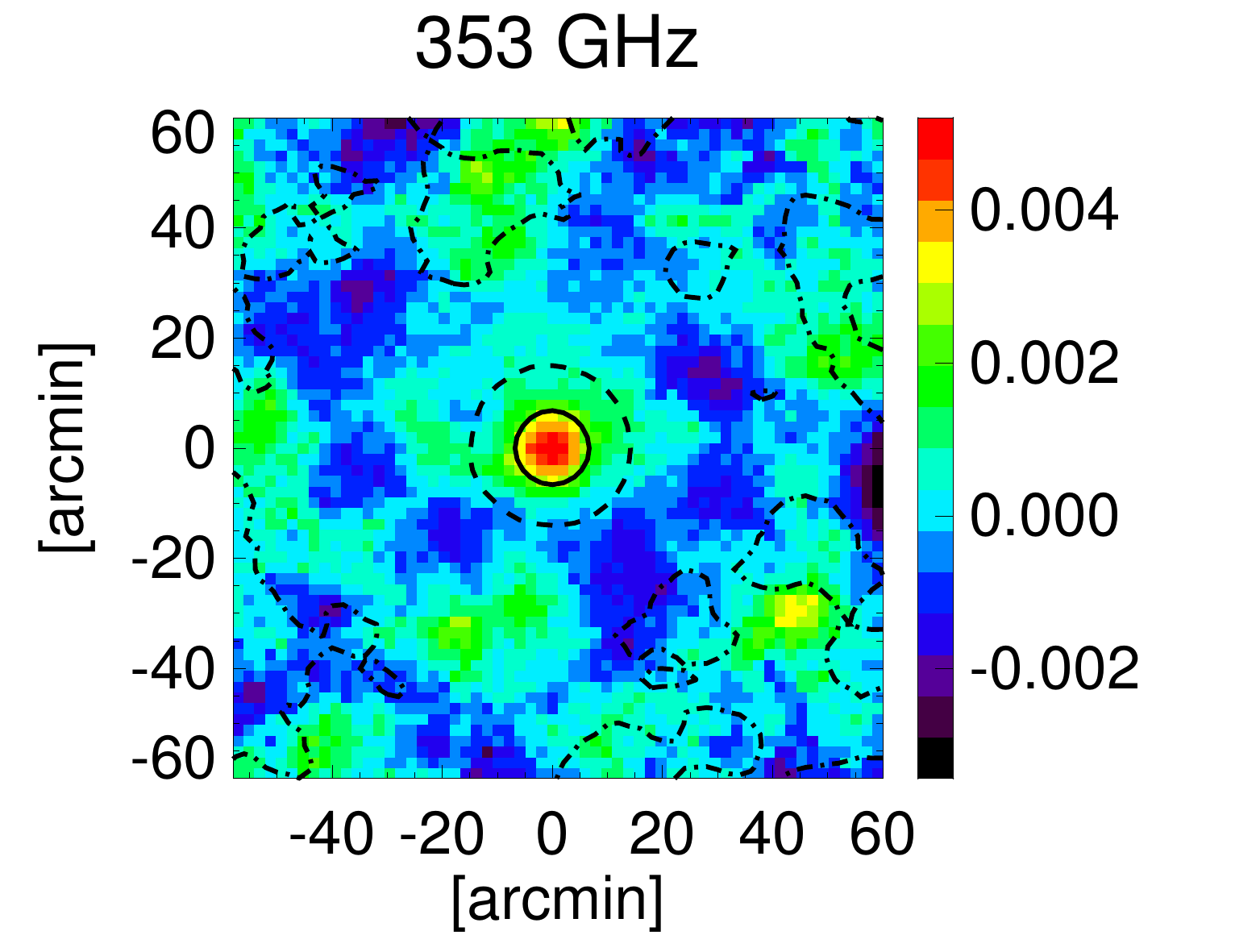}
    \includegraphics[width=0.33\textwidth]{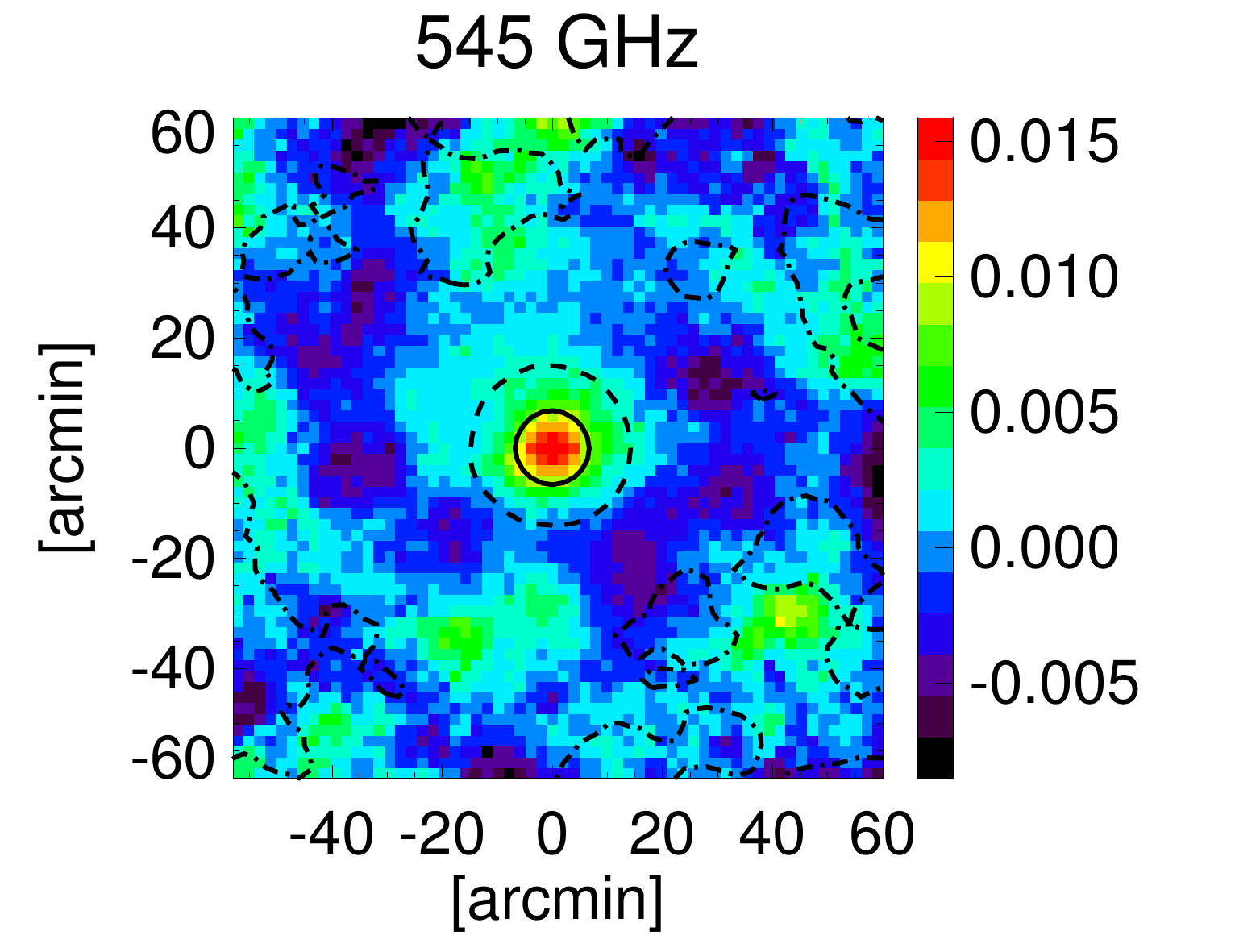}
    \includegraphics[width=0.33\textwidth]{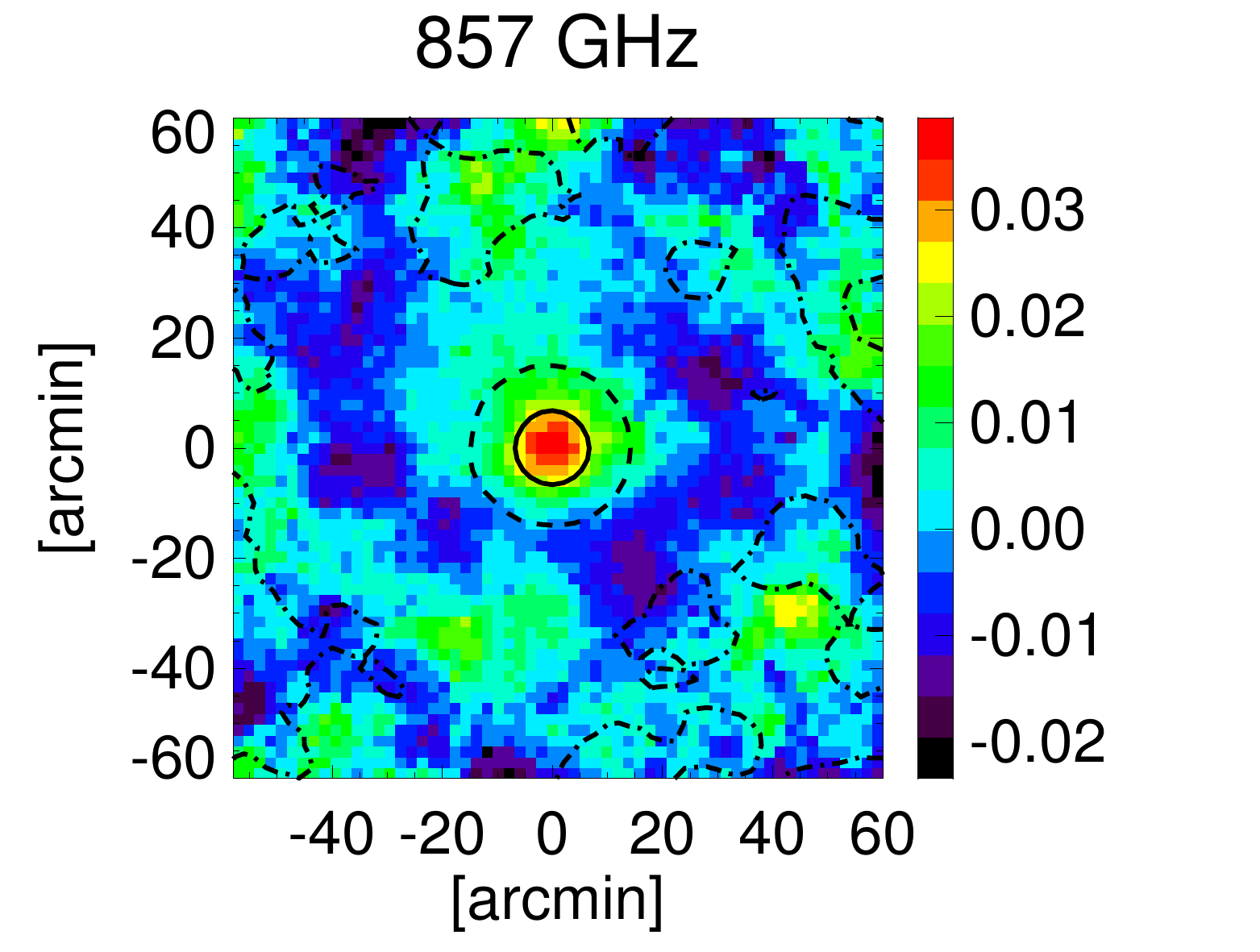}
    \includegraphics[width=0.33\textwidth]{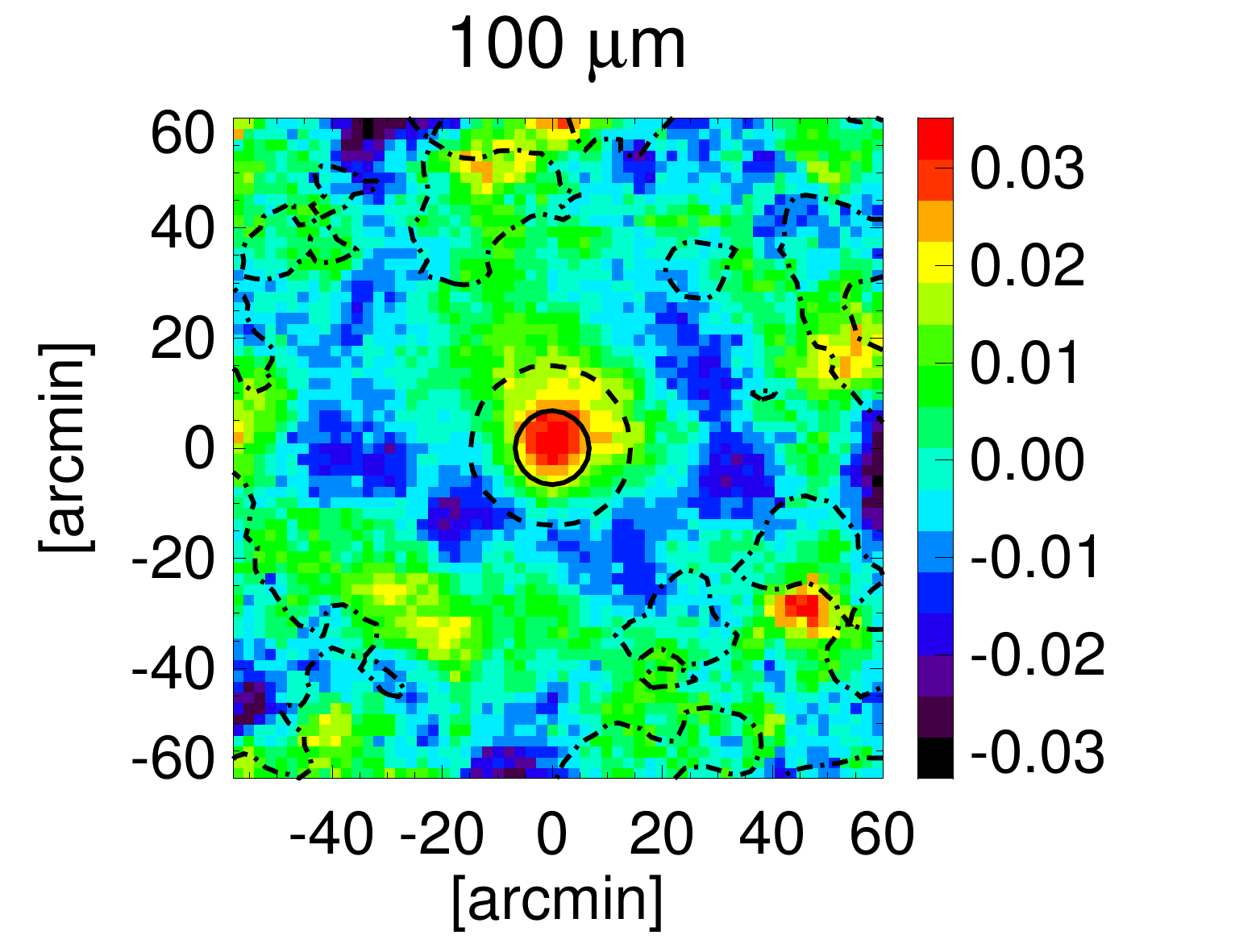}
    \includegraphics[width=0.33\textwidth]{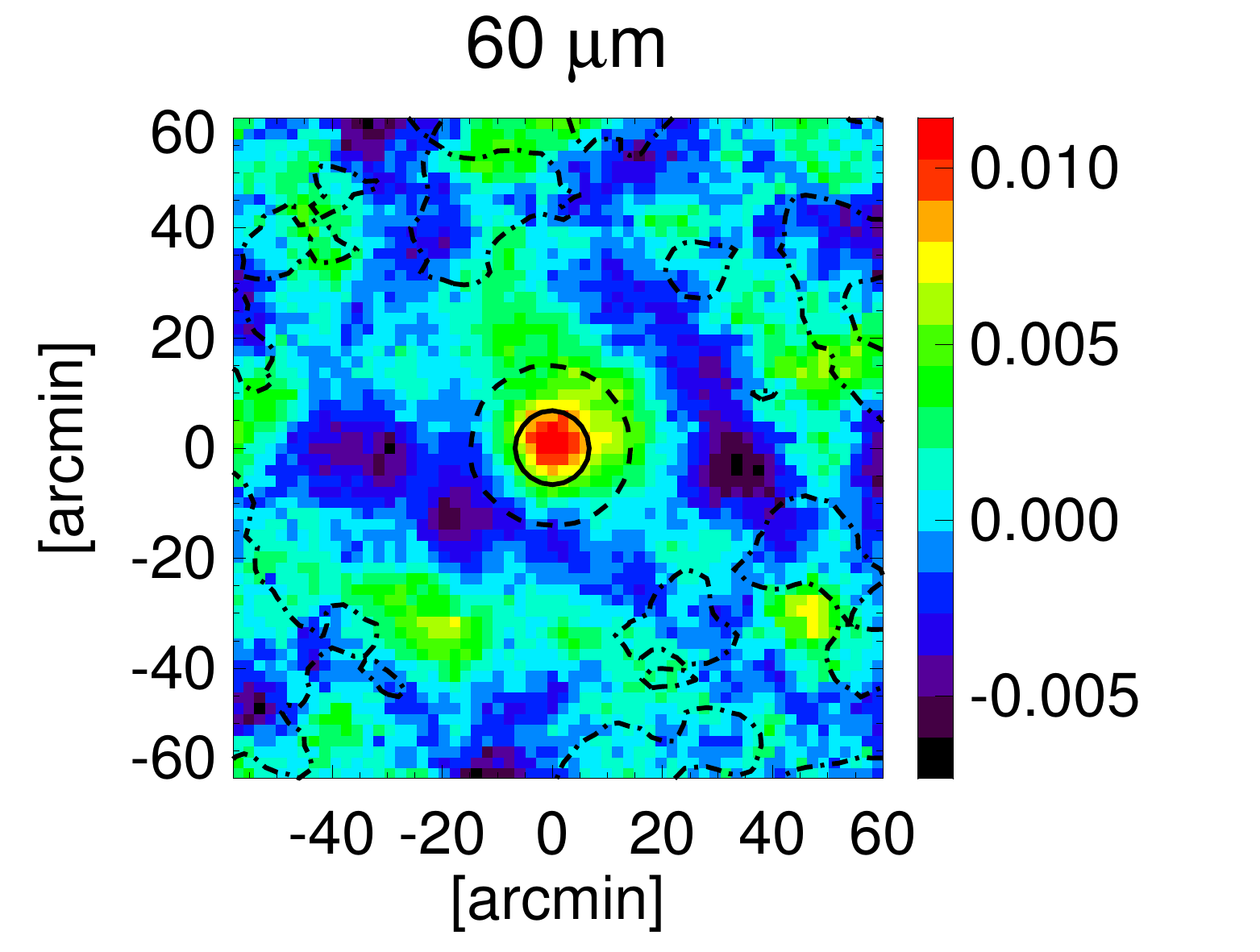}
    \caption{Background- and foreground-cleaned stacked maps, for the final
sample of 645 clusters. The units here are ${\rm MJy}\,{\rm sr}^{-1}$. The
extent of the stacked $y$-signal is represented by black contours for regions
of 0.5, 0.1, and 0 times the maximum of the signal (in solid, dashed, and
dot-dashed lines, respectively). See Sect.~\ref{selected} for further details.}
\label{clean}
\end{figure*}

\subsection{Stacking of the selected fields}\label{selected}
We give the same weight to all the cluster fields, since having a particular
cluster dominate the average signal is not useful for the purposes of examining
the properties of the whole population.
The choice of using a constant patch size is motivated by the fact that the
differences in cluster angular sizes are negligible when dealing with a
10$\arcm$ resolution map. Even if a few tens of very low-$z$ clusters
(say $z < 0.1$) are present within the final sample, the average typical size
of the sample clusters ($\theta_{500}=7\parcm4\pm5\parcm3$, with a median of
5\parcm7) allows us to integrate the total flux in the stacked maps within a
fixed radius, which we choose to be 15$\arcm$. The suitability
of this choice is verified by looking at the integrated signal as a
function of aperture radius.

In Fig.~\ref{noclean} we show the stacked $2\deg\times2\deg$ patches, for the
\Planck-HFI frequencies.  Since no foreground or offset removal has been
performed the dust signature is not easily apparent, but instead we can see
the negative tSZ signal dominating at $\nu\leq143\,$GHz and CMB anisotropies at
217\,GHz, with the dust contribution becoming stronger at $\nu\geq353\,$GHz.
We also stack the CMB and tSZ maps (Sect.~\ref{cmb_sz}), i.e.,
$\Sigma_i M_{\rm CMB}^i$ and $\Sigma_i M_{\rm tSZ}^i$, summing over all
the selected clusters $i$. These two quantities are then subtracted from the
raw frequency maps ($\Sigma_i M_{\nu}^i$ shown in Fig.~\ref{noclean}),
specifically calculating $M'_{\nu}=\big(\Sigma_i M_{\nu}^i
 - \Sigma_i M_{\rm CMB}^i - \Sigma_i f_{{\rm tSZ},\nu}M_{\rm SZ}^i \big)/N$,
where $f_{{\rm tSZ},\nu}$ is the conversion factor from the tSZ Compton
parameter $y$ to ${\rm MJy}\,{\rm sr}^{-1}$ for each frequency $\nu$.
Following \cite{GiardMontier2008}, we then perform a background-subtraction
procedure by fitting a 3rd-order polynomial surface to the map region for which
the cluster-centric distance is above 10\arcm. Finally we also subtract the
average signal found for pixels with a distance from the centre that lies
between 0\pdeg5 and $1\deg$, which is used to compute the zero level of the
map.

\begin{figure*}
  \centering
    \includegraphics[width=0.33\textwidth]{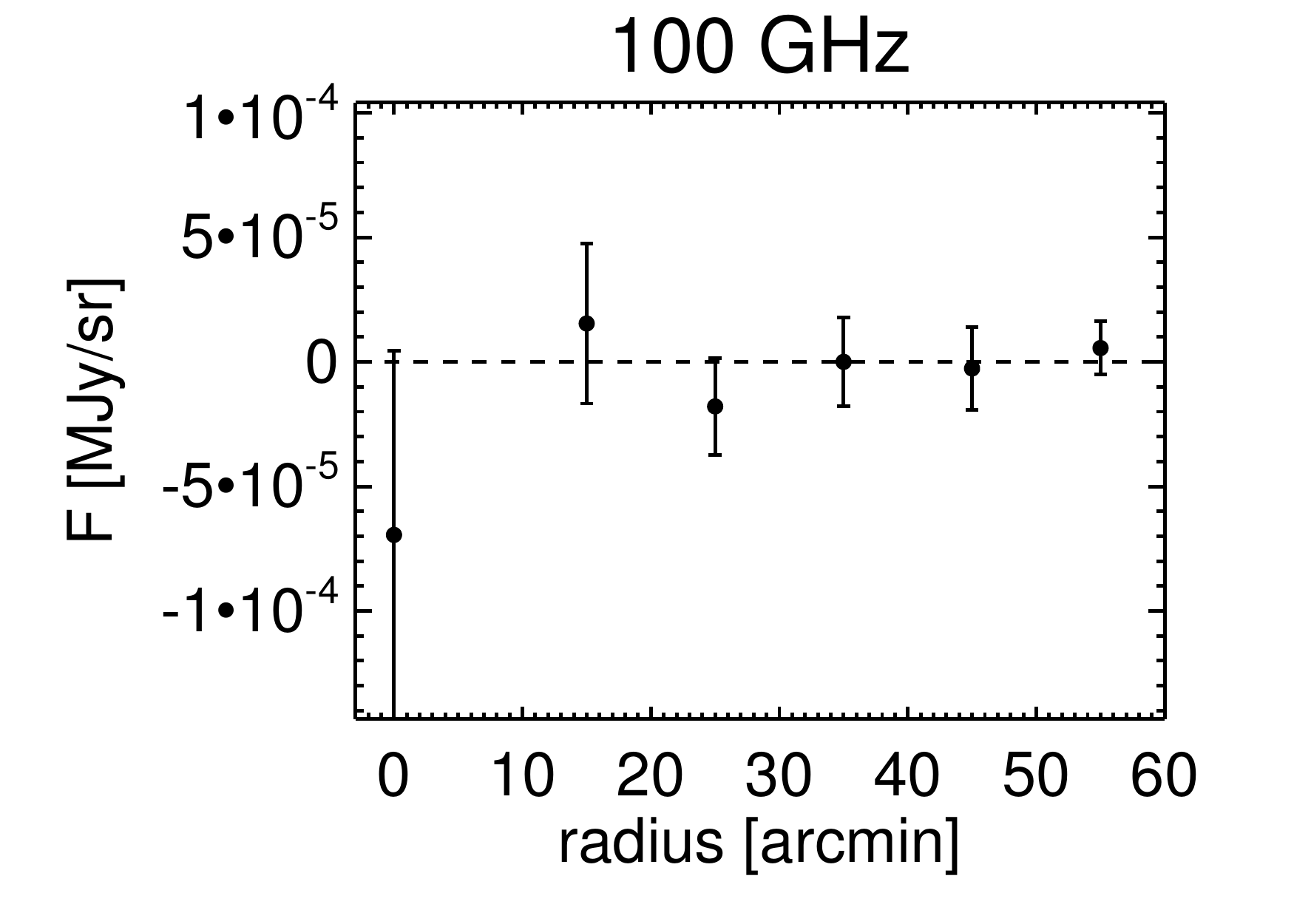}
    \includegraphics[width=0.33\textwidth]{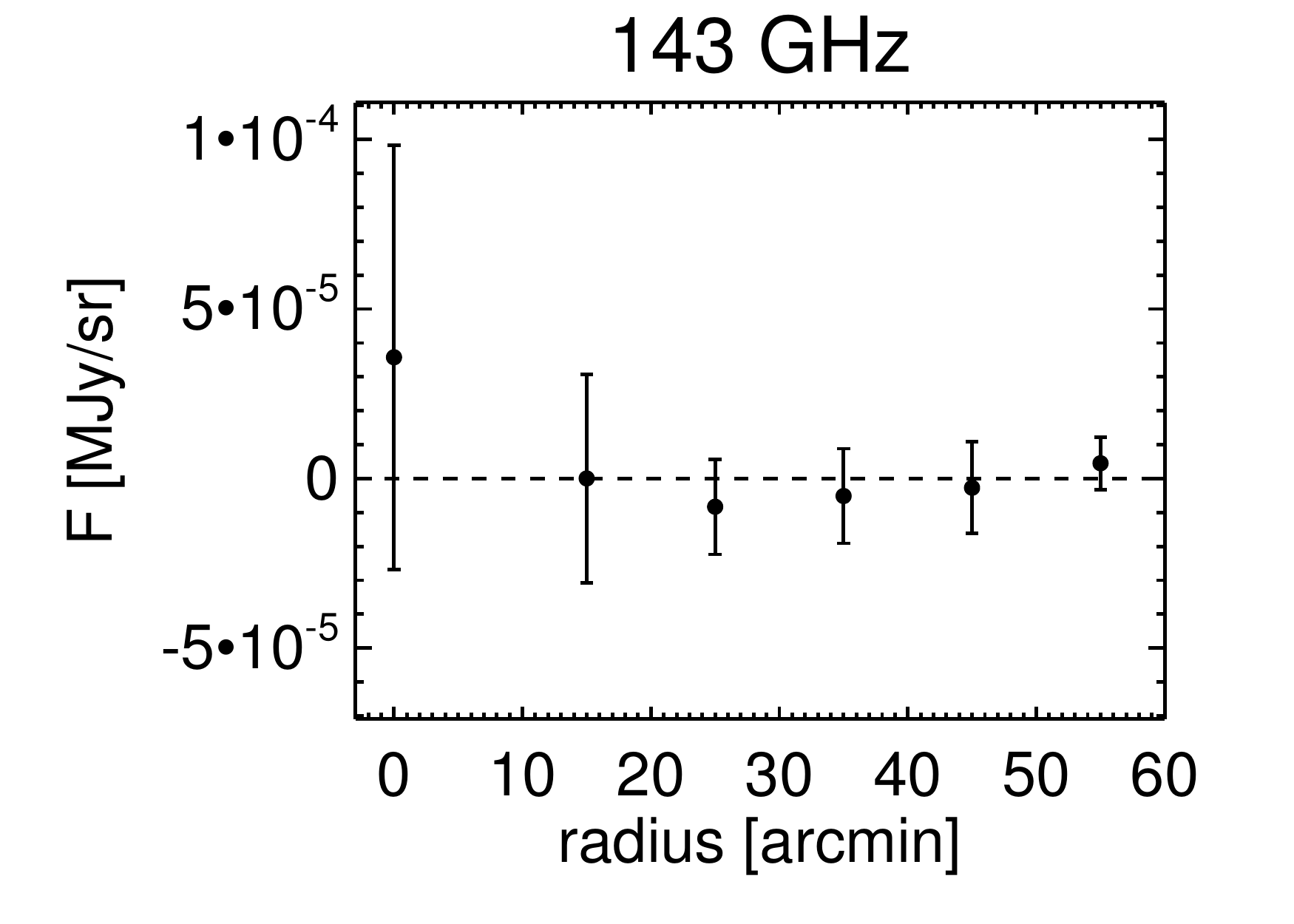}
    \includegraphics[width=0.33\textwidth]{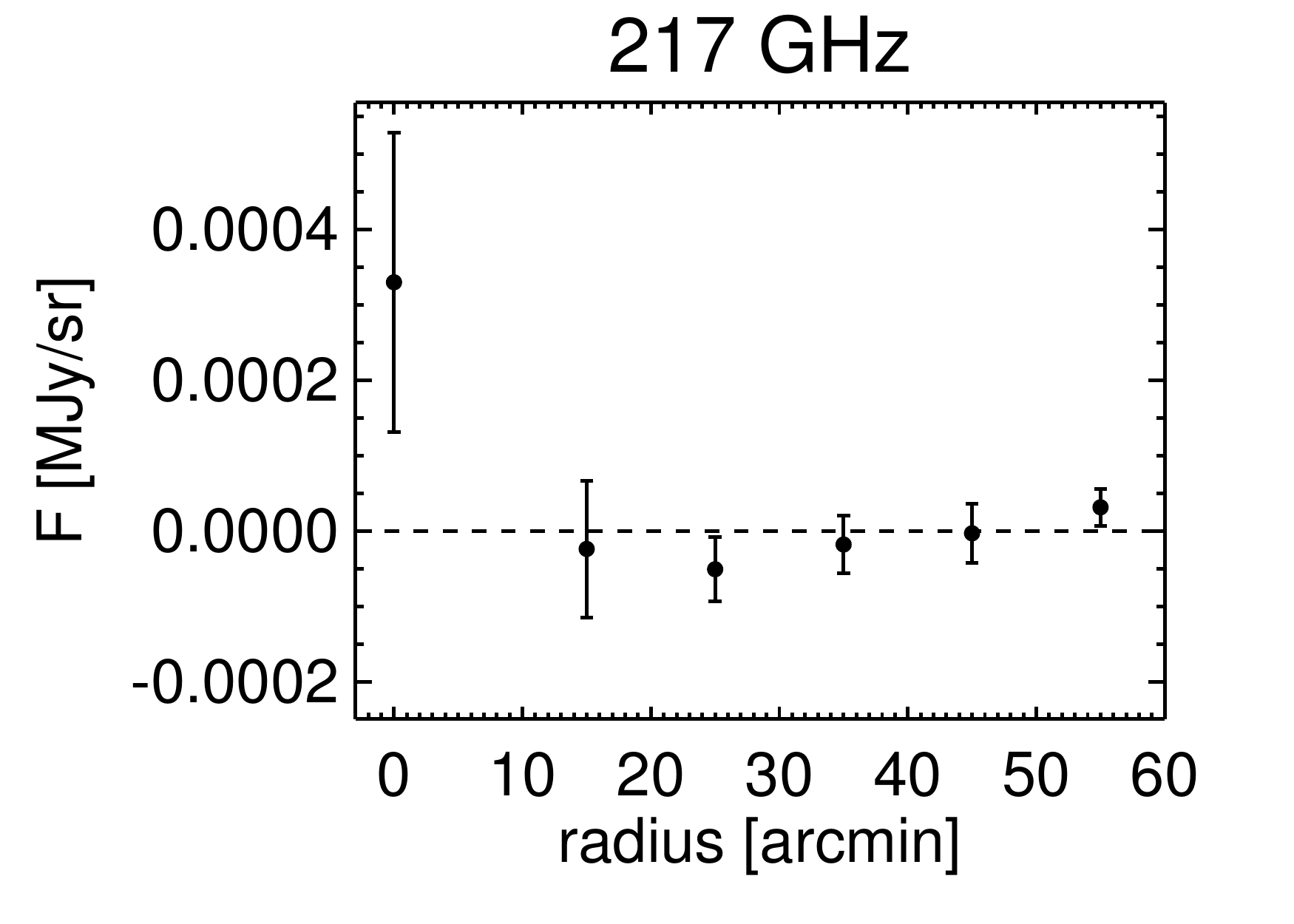}
    \includegraphics[width=0.33\textwidth]{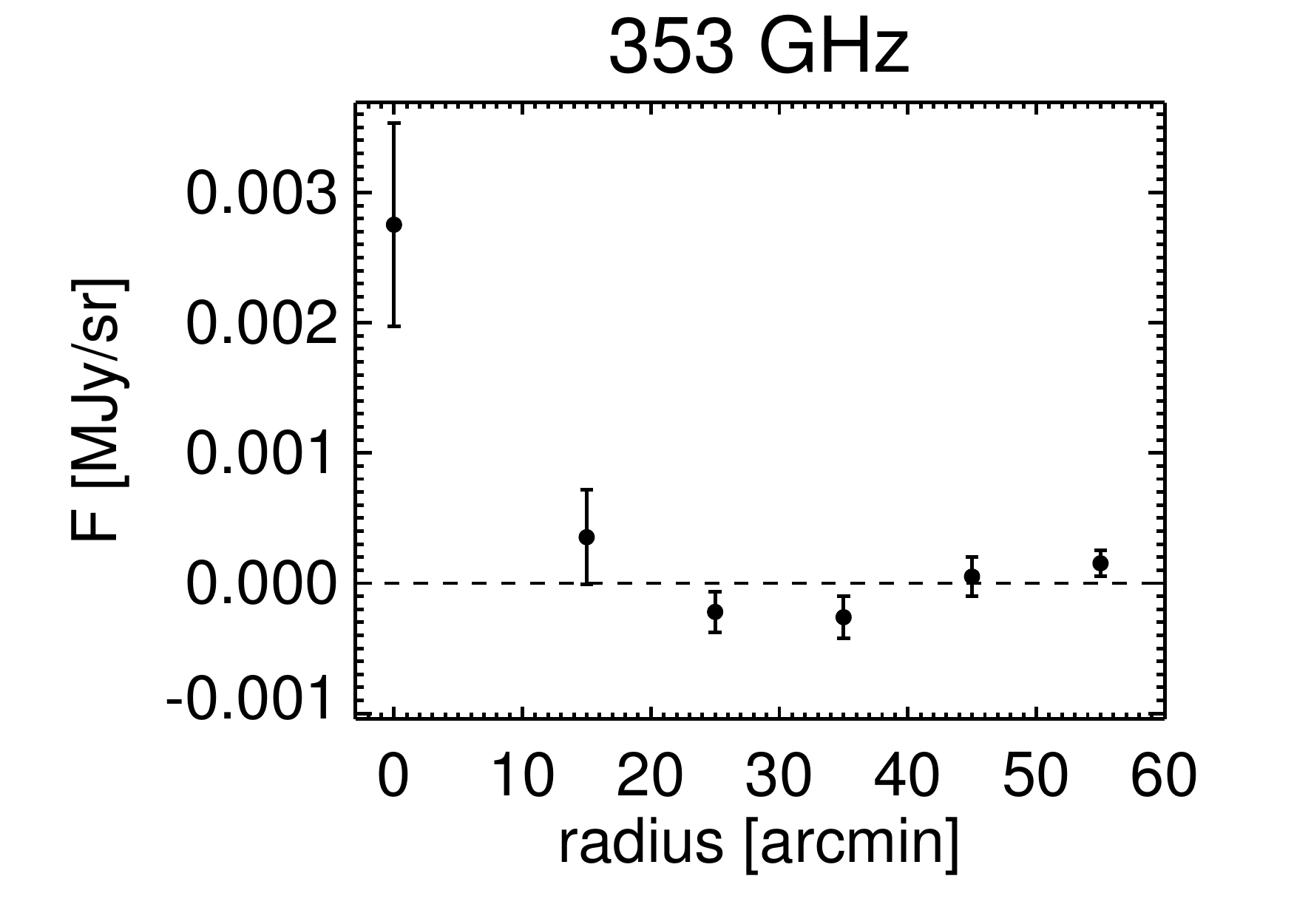}
    \includegraphics[width=0.33\textwidth]{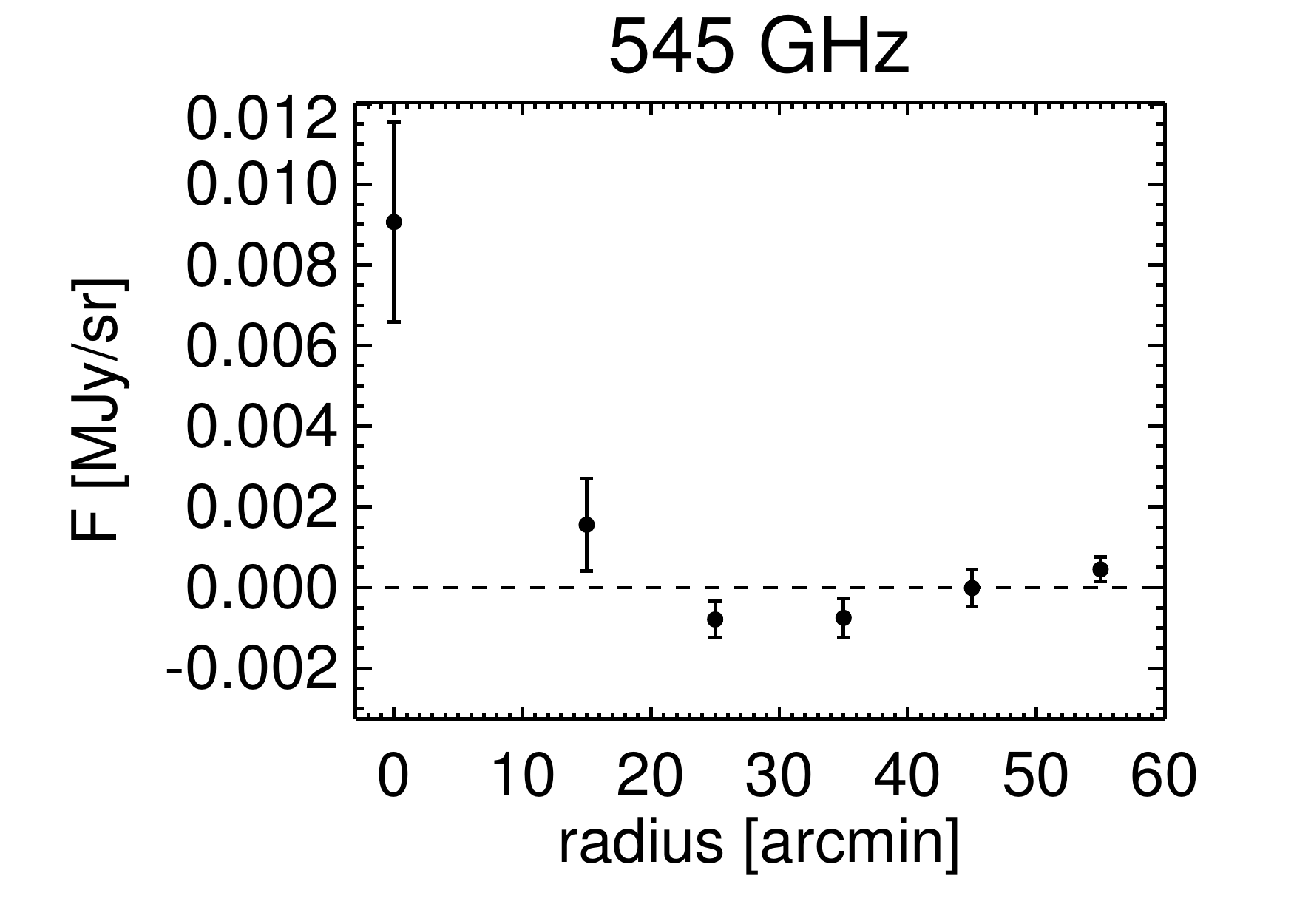}
    \includegraphics[width=0.33\textwidth]{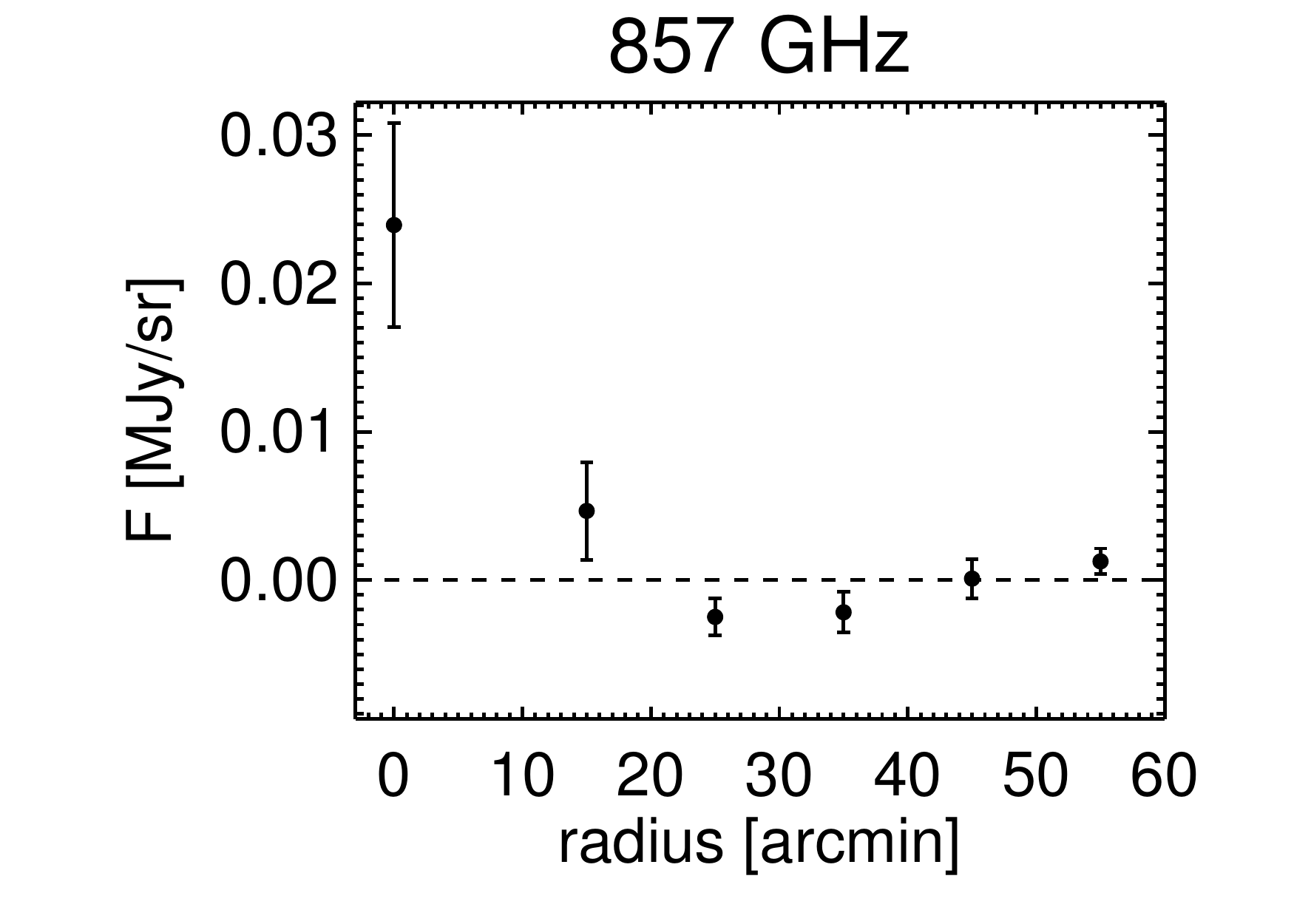}
    \includegraphics[width=0.33\textwidth]{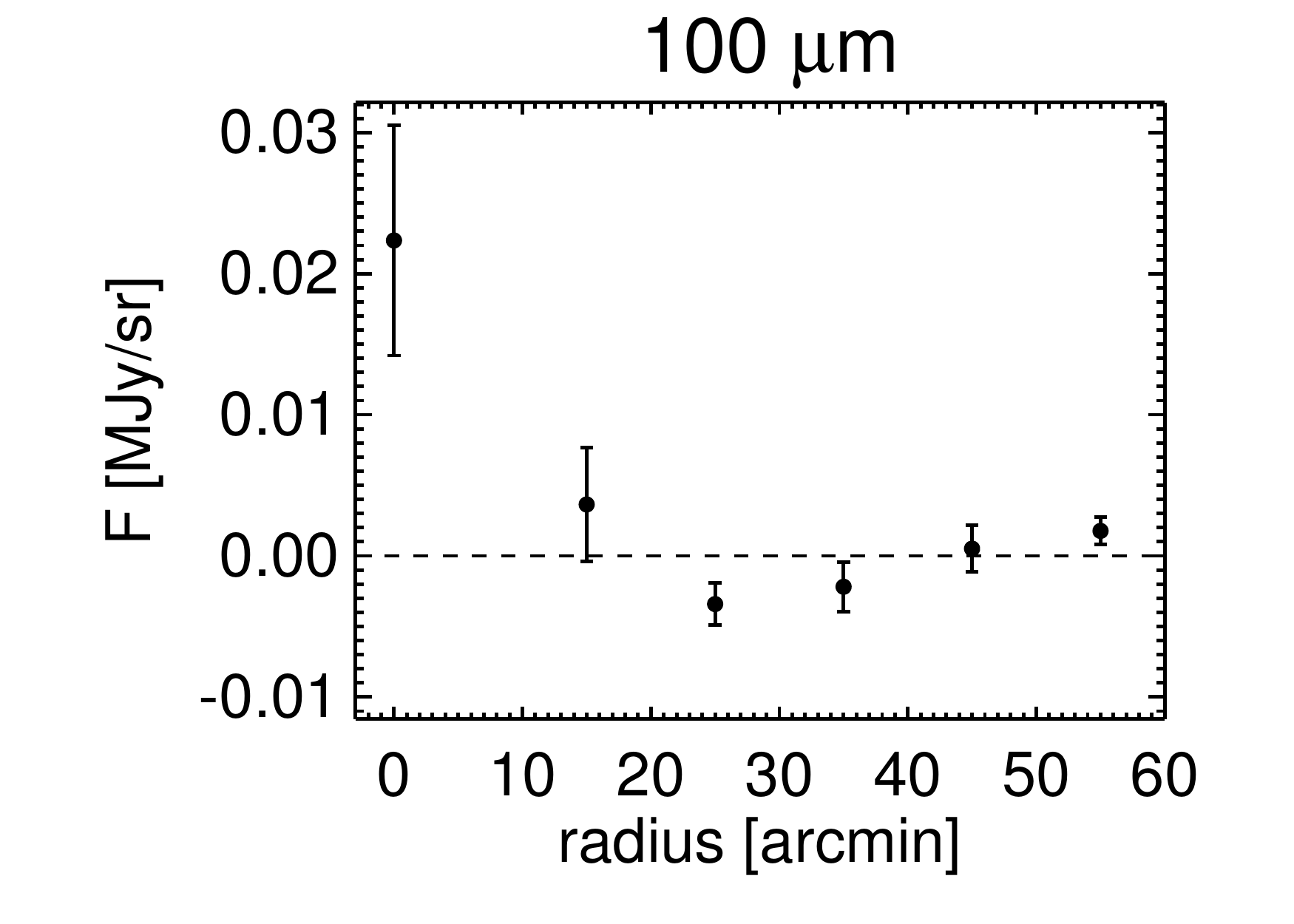}
    \includegraphics[width=0.33\textwidth]{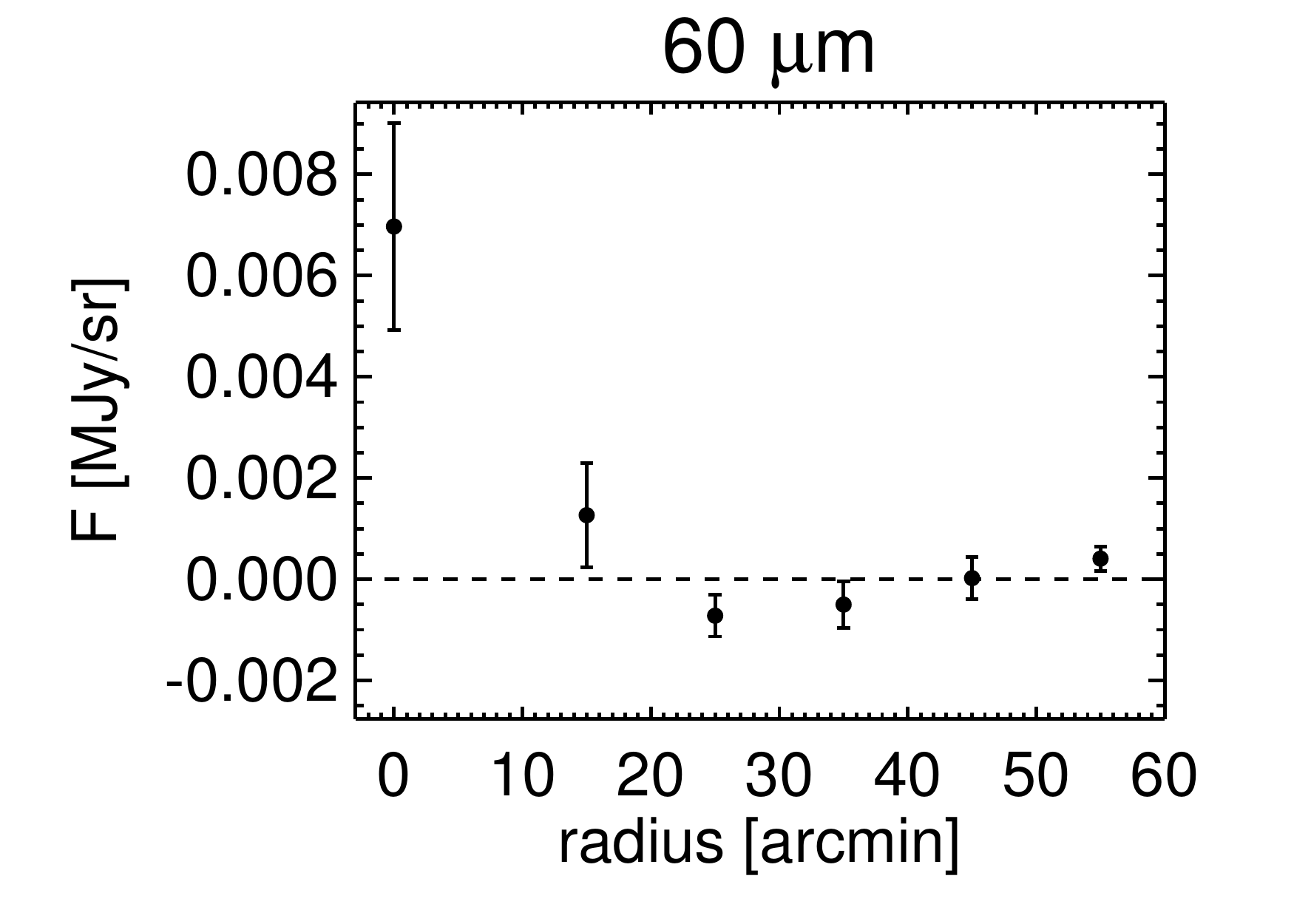}
      \caption{Radial profiles obtained from the background- and
foreground-cleaned stacked maps, for the sample of 645 clusters. The units
here are ${\rm MJy}\,{\rm sr}^{-1}$. The black points correspond to the values
obtained as the average of the pixels contained within each region considered
and associated uncertainties have been obtained using bootstrap resampling.}
         \label{radial_profiles}
\end{figure*}

The cleaned and stacked maps are shown in Fig.~\ref{clean}. The same results
are obtained if foregrounds and offsets are subtracted cluster by cluster or if
we directly stack the cleaned cluster-centred patches. Although we already
have hints of a signal at the centre of 143\,GHz map, the detection of a
significant central positive peak starts at 217\,GHz, and is very clearly
observed at $\nu\geq353$\,GHz. Between 100 and 217\,GHz the signal is expected
to be fainter, according to the typical SED of thermal dust emission. The black
contours in Fig.~\ref{clean} allow comparison with the distribution of the
cluster hot gas, representing where the stacked $y$-map has a value that is
0.5, 0.1, and 0 times its maximum. As expected, the recovered IR emission
follows the distribution of the main cluster baryonic component, and thus the
extent of the clusters. Fig.~\ref{radial_profiles} represents the average
intensity profiles as a function of radius. 
 
Since IR dust emission cannot be detected for individual clusters, average
values have been obtained, and the associated uncertainties determined
using a bootstrap approach.  This consists of constructing and stacking
many (300) cluster samples obtained by randomly replacing sources from the
original sample, so that each of them contains the same number of clusters
as the initial one. The statistical properties of the population being stacked
can be then determined by looking at the mean and standard deviation of the
flux found in the stacked maps corresponding to each of the resampled cluster
lists.  We checked that the average values obtained with this resampling
technique are equal to what we obtained directly on the original stacked map
(without any resampling). This is important, since it indicates that we are
indeed stacking a homogeneous population of objects, and that the detected
signal is not due to only a small number of a clusters.
The mean values recovered are also consistent with the expectations described
in \citet{MontierGiard2005}, given the redshift distribution of the sample
considered here.
 
Figure ~\ref{radial_profiles} shows that we find no significant detection at
100 and 143\,GHz, while the detection starts to become strong at
$\nu\geq353\,$GHz, consistent with what we saw in Fig.~\ref{clean}.  In
Table~\ref{table:sed} we report, for each frequency: the average fluxes, $F$,
found when integrating out to 15$\arcm$ from the centre; the standard deviation
found using the bootstrap resampling, $\Delta F_{\rm b}$; and an estimate of the
uncertainty on the flux at each frequency, $\Delta F$. The flux uncertainties,
$\Delta F$, have been obtained as the standard deviation of the fluxes found
at random positions in a $2\deg\times2\deg$ region, located further than
15\arcm\ from the centre, both using the cluster-centred stacked maps and the
``depointed'' stacked maps. The latter correspond to regions centred $1\deg$
away from the cluster Galactic latitude and/or longitude. They are also used
to test our detection against stacking artefacts, as will be discussed in
Sect.~\ref{tests}.

\subsection{Robustness tests} \label{tests}
\begin{figure}
 \centering
   \includegraphics[width=0.32\columnwidth]{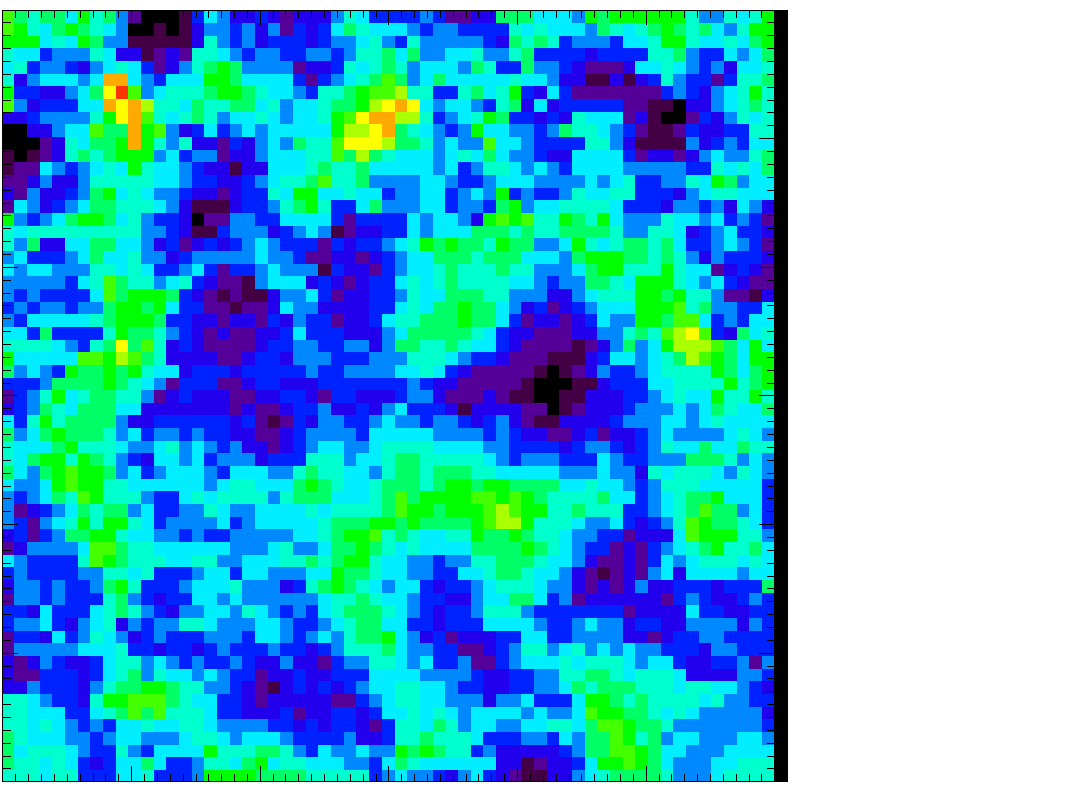}
   \includegraphics[width=0.32\columnwidth]{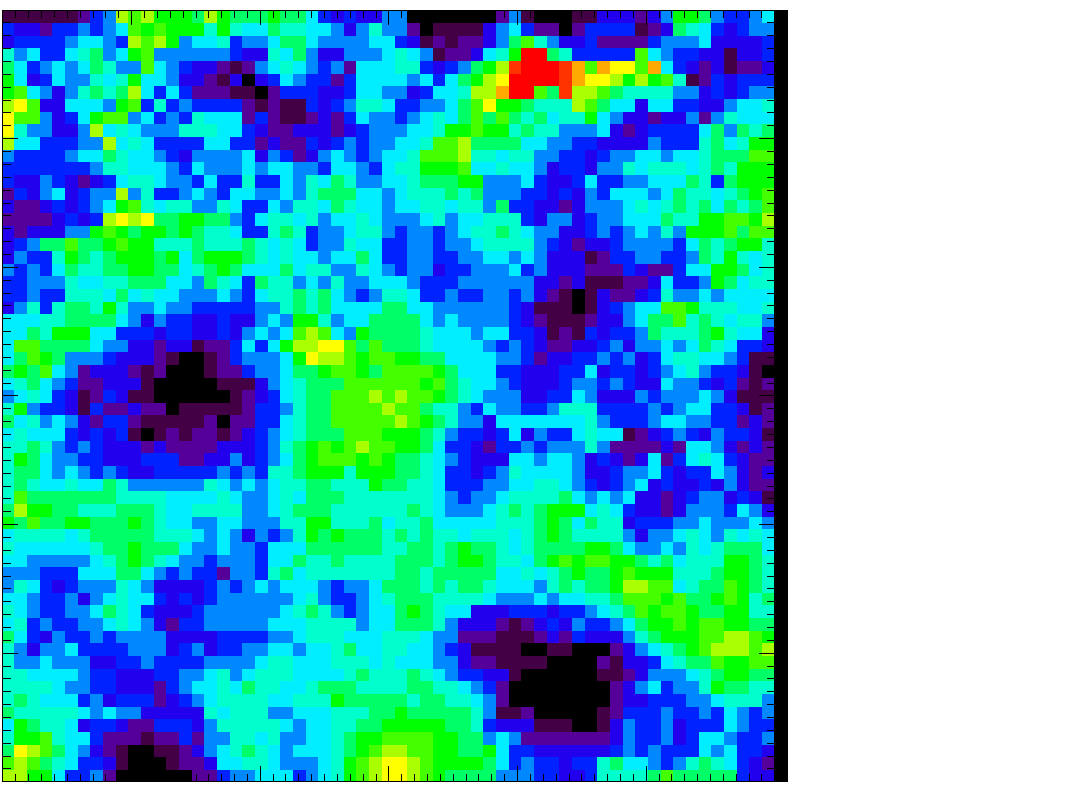}
   \includegraphics[width=0.32\columnwidth]{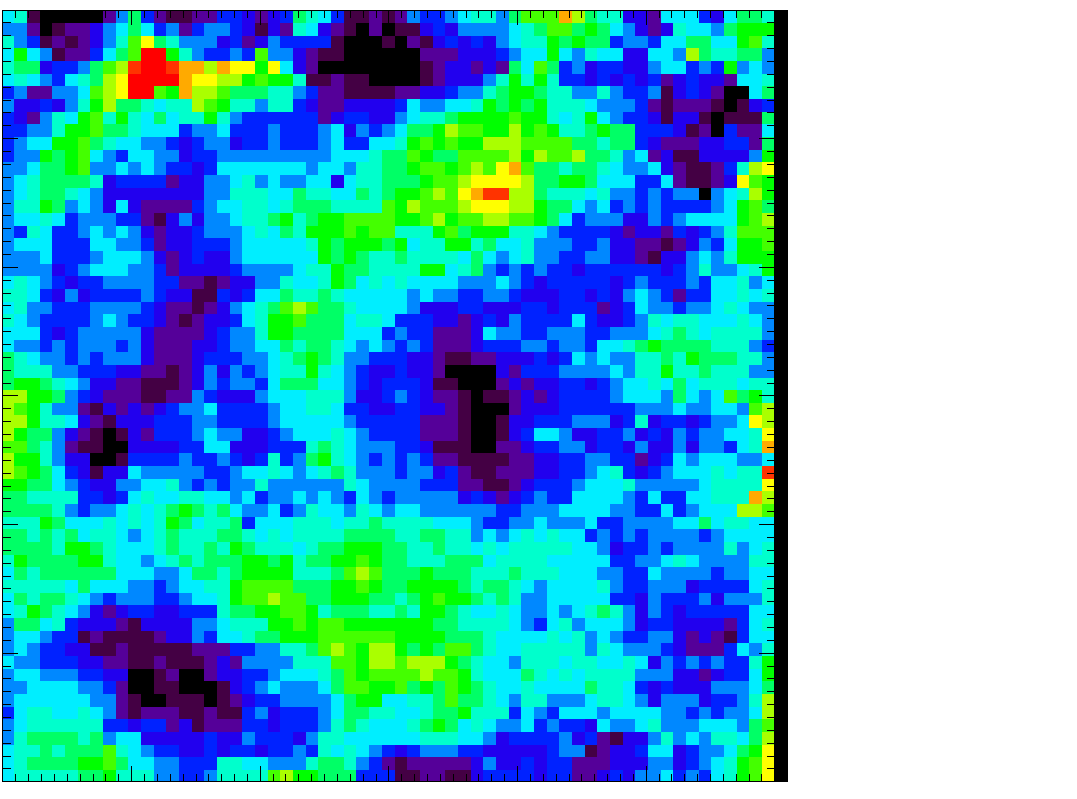}
   \includegraphics[width=0.32\columnwidth]{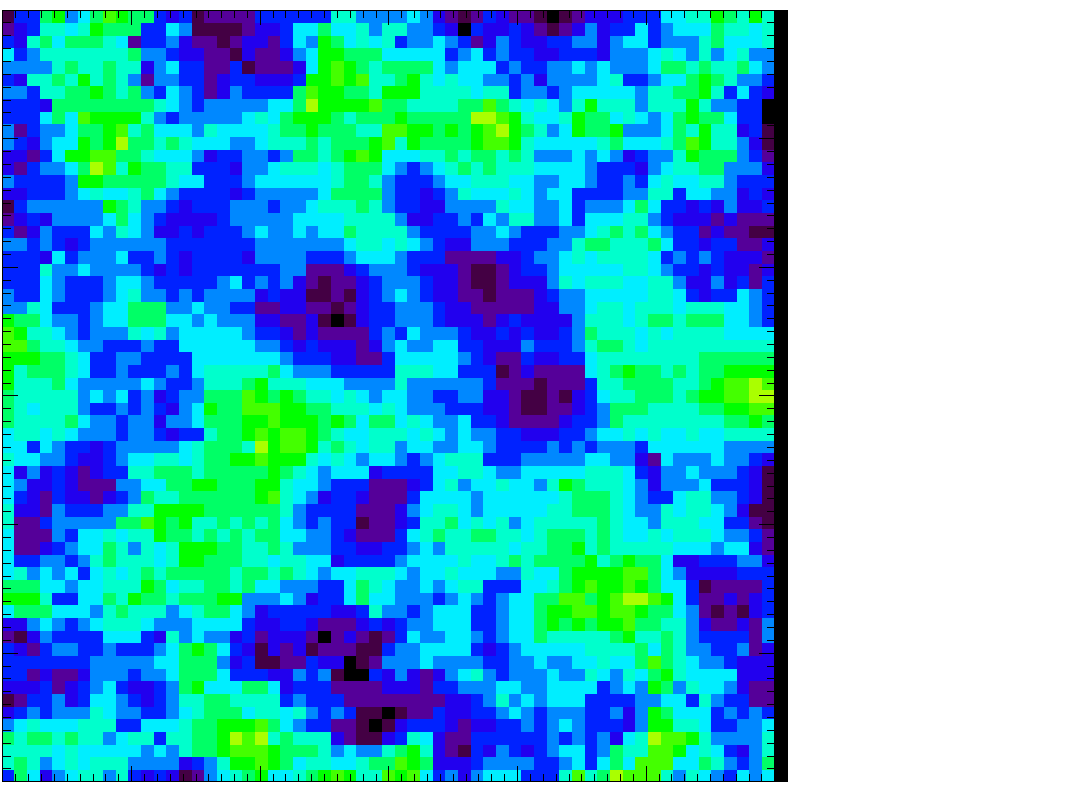}
   \includegraphics[width=0.32\columnwidth]{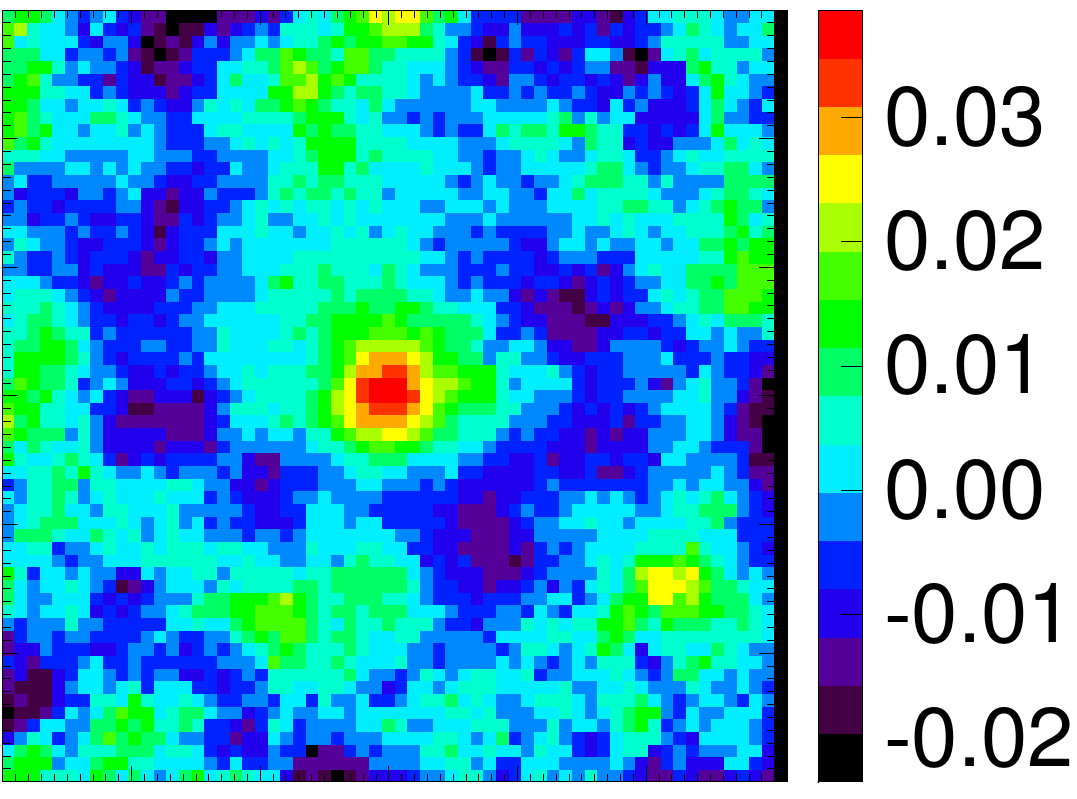}
   \includegraphics[width=0.32\columnwidth]{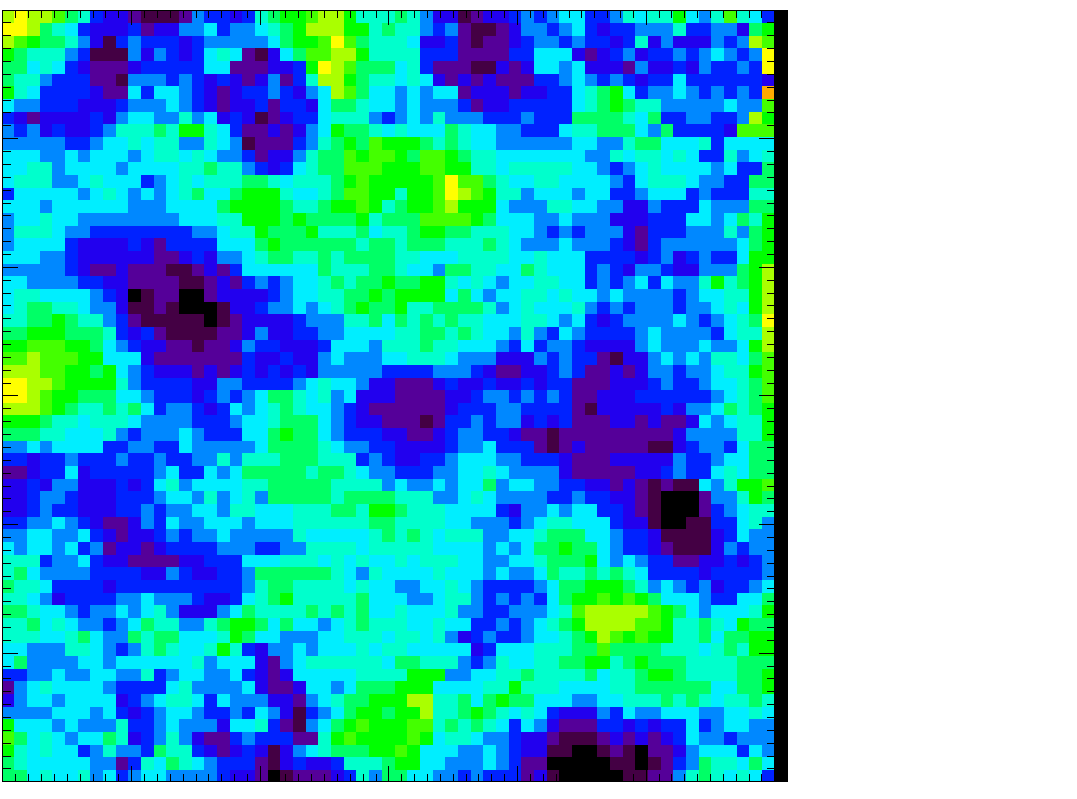}
   \includegraphics[width=0.32\columnwidth]{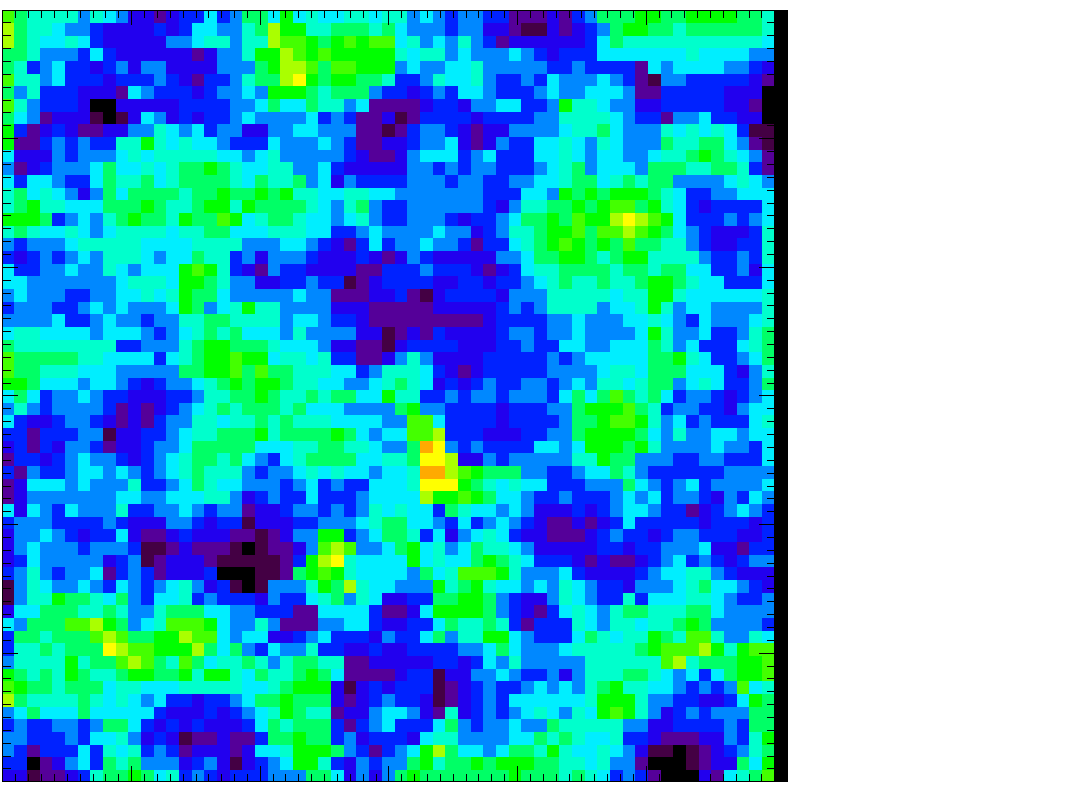}
   \includegraphics[width=0.32\columnwidth]{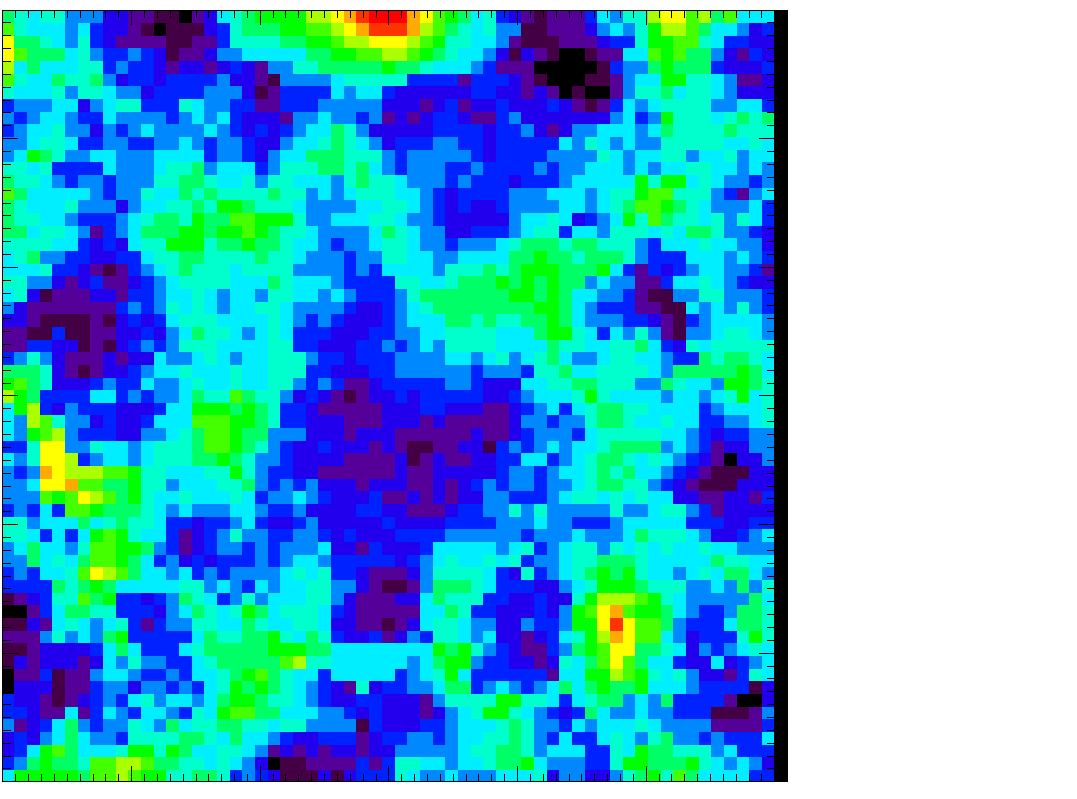}
   \includegraphics[width=0.32\columnwidth]{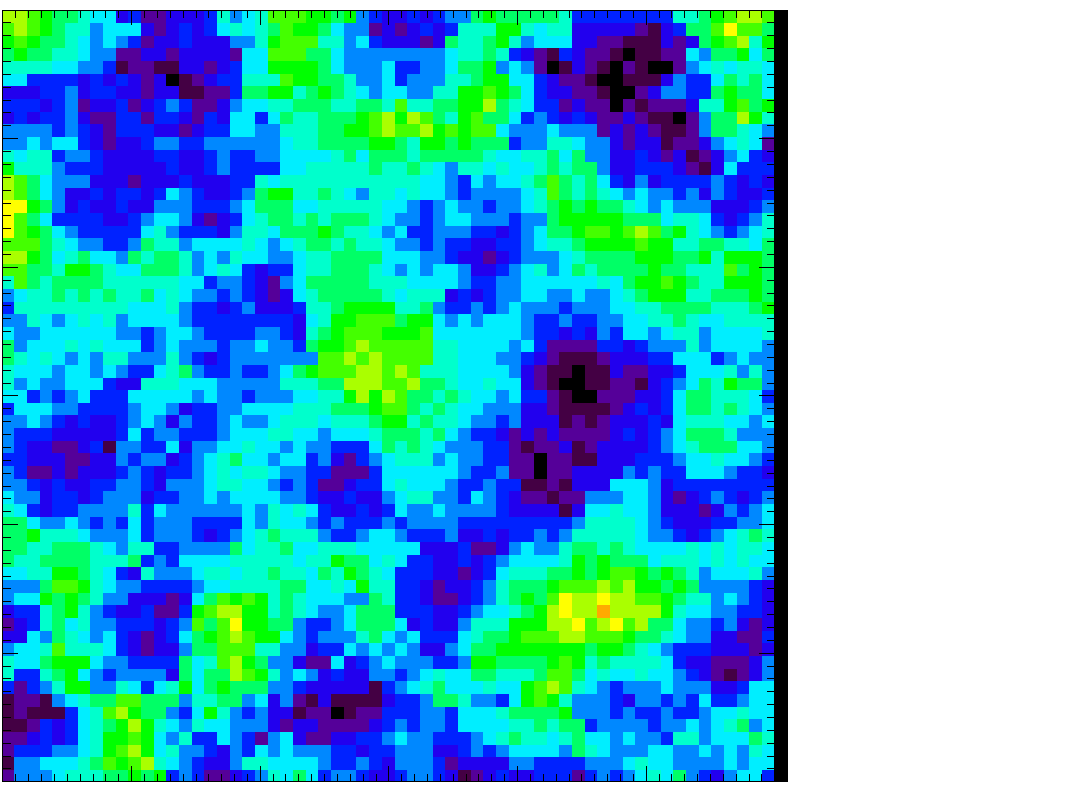}   
    \caption{Tests of stacking on ``depointed'' regions.  The central panel
shows the stacked result obtained at 857\,GHz for the 645 cluster-centred
patches (as in Fig.~\ref{clean}).  This is surrounded by the eight neighboring
maps obtained by changing the cluster Galactic longitude or latitude (or both)
by $\pm1\deg$.}
        \label{depointing}
\end{figure}

In order to test the robustness of our results, we have performed various
checks, following an approach similar to that of \cite{MontierGiard2005}.
Figure~\ref{depointing} shows the $2\deg\times2\deg$ depointed maps that we
obtain at 857\,GHz when we repeat the same stacking procedure by changing the
cluster Galactic longitudes and/or latitudes by $\pm1\deg$. This has been done
for all the frequency channels and shows that the detection is not an artefact
of the adopted stacking scheme. The mean of the fluxes obtained at the centre
of the depointed regions is consistent with zero within the uncertainties
(i.e., $\Delta F$).

The approach adopted to derive the uncertainty on the flux $\Delta F$ has also
been applied to a random sample of positions on the sky, whose Galactic
latitude distribution follows that of the real clusters in our sample.
The derived uncertainties, $\Delta F_{\rm ran}$, are listed in
Table~\ref{table:sed}.  As for the depointed regions, the mean fluxes obtained
at the centre of the random patches are consistent with zero, within the given
uncertainties. The values obtained for $\Delta F_{\rm ran}$ are systematically
higher than $\Delta F$. This was to be expected, since \Planck\ blind tSZ
detections are more likely in regions that are cleaner of dust contamination.
For different cuts in Galactic latitude, moving away from the Galactic plane,
$\Delta F_{\rm ran}$ decreases. This might indicate that some residual
contamination due to Galactic dust emission is present. Indeed, in
Fig.~\ref{clean} we can see a correlation between frequencies for the residual
fluctuations in the region surrounding the clusters.  Such a correlation
between frequencies could be also introduced by the process of subtracting the
contamination maps (CMB and tSZ), since these are built from the same
\Planck-HFI maps. However the CMB anisotropies and the tSZ signal are both
negligible at $\nu\geq545\,$GHz. The uncertainty $\Delta F_{\rm ran}$ is of
the order as $\Delta F_{\rm b}$ and $\Delta F$ at the frequencies for which we
have a significant detection; hence this contribution does not dominate the
signal and we can consider it to be accounted for in the error budget.  For
this reason, we do not impose any extra selection cut in Galactic latitude in
order to maximize the sample size.
  
As a further cross-check, we have tested the robustness of the results by
alternatively adding and subtracting the patches centred at the cluster
positions. This approach shows that none of the individual patches dominates
the final average signal, in agreement with the results of the bootstrap
resampling procedure.

%% file: 04_results.tex
\subsection{Assumed SED shape}\label{SEDshape}
\begin{figure}
  \centering
    \includegraphics[width=\columnwidth]{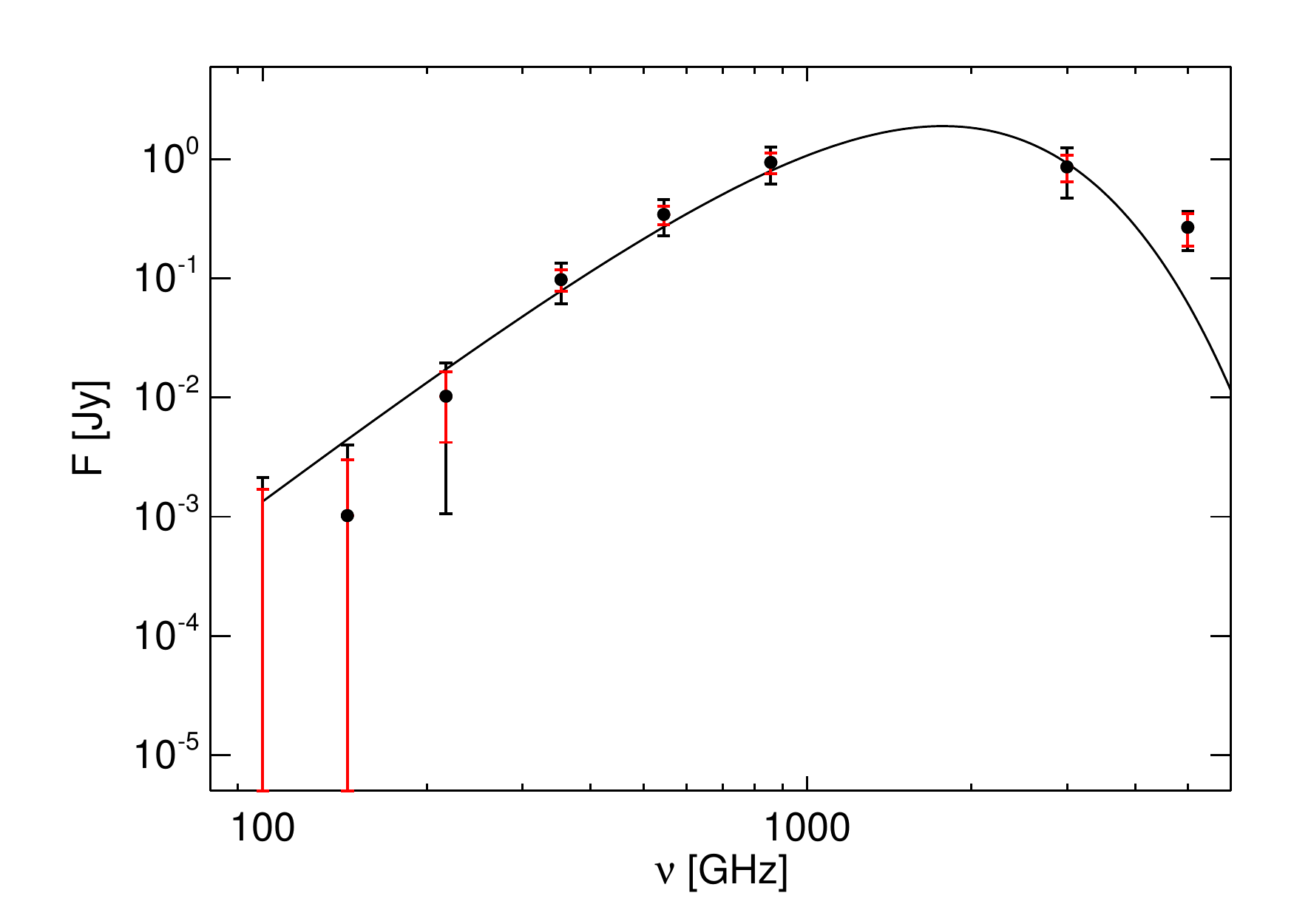}
      \caption{Average SED for galaxy clusters.  In black points we show the
flux, as a function of frequency, found in the co-added maps obtained for a
sample of 645 clusters. The error bars in black correspond to the dispersion
estimated using bootstrap resampling ($\Delta F_{\rm b}$,
Table~\ref{table:sed}). The red error bars have been estimated as the standard
deviation of the flux integrated at random positions away from each cluster
region ($\Delta F$, Table~\ref{table:sed}). The black solid line shows the
best-fit modified blackbody model (with $\beta=1.5$). Note that the highest
frequency point ($60\,\mu$m from IRAS) is not used in the fit.}
         \label{fig:sed}
\end{figure}

\begin{figure*}
  \centering
    \includegraphics[width=0.49\textwidth]{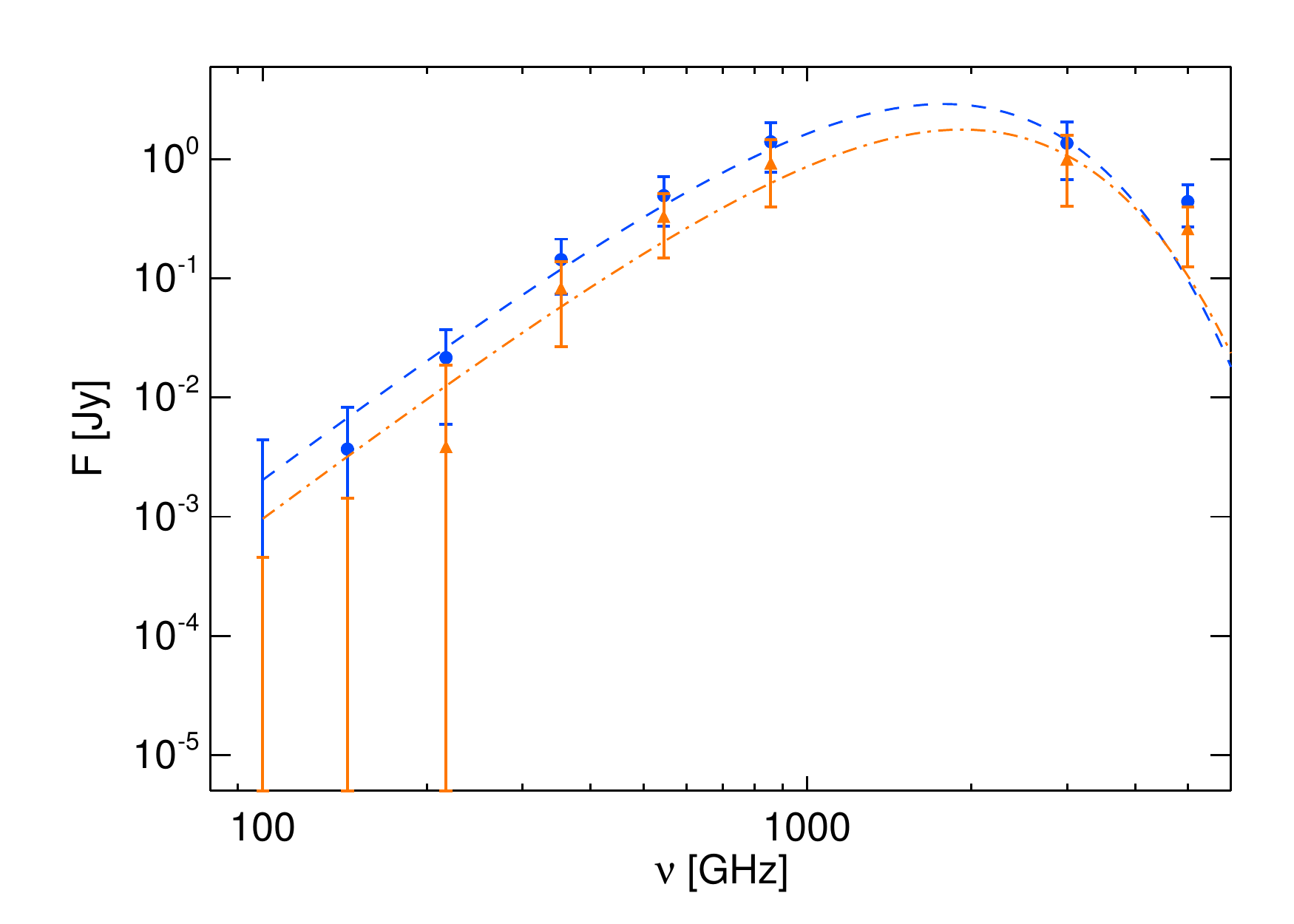}
    \includegraphics[width=0.49\textwidth]{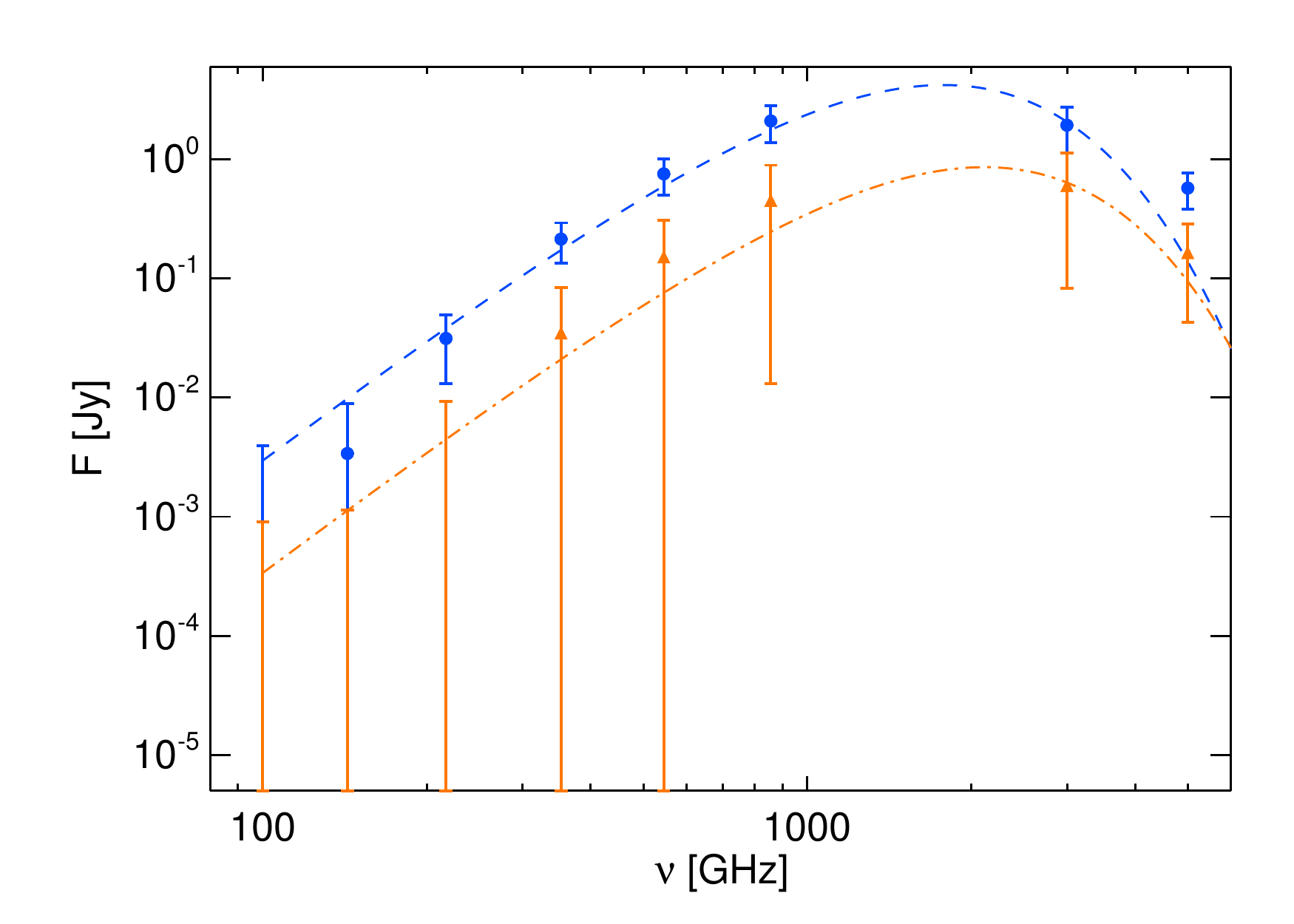}
    \caption{SEDs for cluster sub-samples.  {\it Left}: In blue circles the
254 clusters at $z>0.25$ and in orange triangles the 307 clusters at
$z\le0.25$ (307 clusters), plotted along with the corresponding best-fit models
(dashed and dash-dotted lines, respectively). {\it Right}: In blue the
241 clusters with $M_{\rm tot}>5.5\times10^{14}{\rm M}_{\odot}$ and in orange
the 320 clusters with $M_{\rm tot}\leq5.5\times10^{14}{\rm M}_{\odot}$, along
with the best-fit models (dashed and dash-dotted lines, respectively).
The fluxes at each frequency are also listed in Tables~\ref{table:sed_z} and
\ref{table:sed_m}. The rightmost (IRAS) point is not used in the fit.}
         \label{subsamples}
\end{figure*}

Using the cleaned stacked maps obtained in Sect.~\ref{sec:stacking}, we can
derive the IR fluxes for each of the frequencies considered. In
Fig.~\ref{fig:sed} we show the average SED of galaxy clusters from
60$\,\mu$m to 3\,mm (as listed in Table~\ref{table:sed}).
We divide our sample into two different sub-samples, according to their
redshifts and their masses.  Specifically, we consider clusters above and
below $z=0.25$ (with 254 and 307 sources, respectively) and above and below
$M^{500}_{\rm tot}=5.5\times10^{14}\,{\rm M}_{\odot}$
(241 and 320 sources, respectively).\footnote{$M^{500}_{\rm tot}$ is the total
mass contained within a radius ($\theta_{500}$) at which the mean cluster
density is 500 times the critical density of the Universe.}
The left and right panels in Fig.~\ref{subsamples} show the measured SED for
the low (orange) and high (blue) redshift and mass subsamples, respectively.
The corresponding fluxes and associated uncertainties are also listed in
Tables~\ref{table:sed_z} and \ref{table:sed_m}. 

\begin{figure}
  \centering
    \includegraphics[width=\columnwidth]{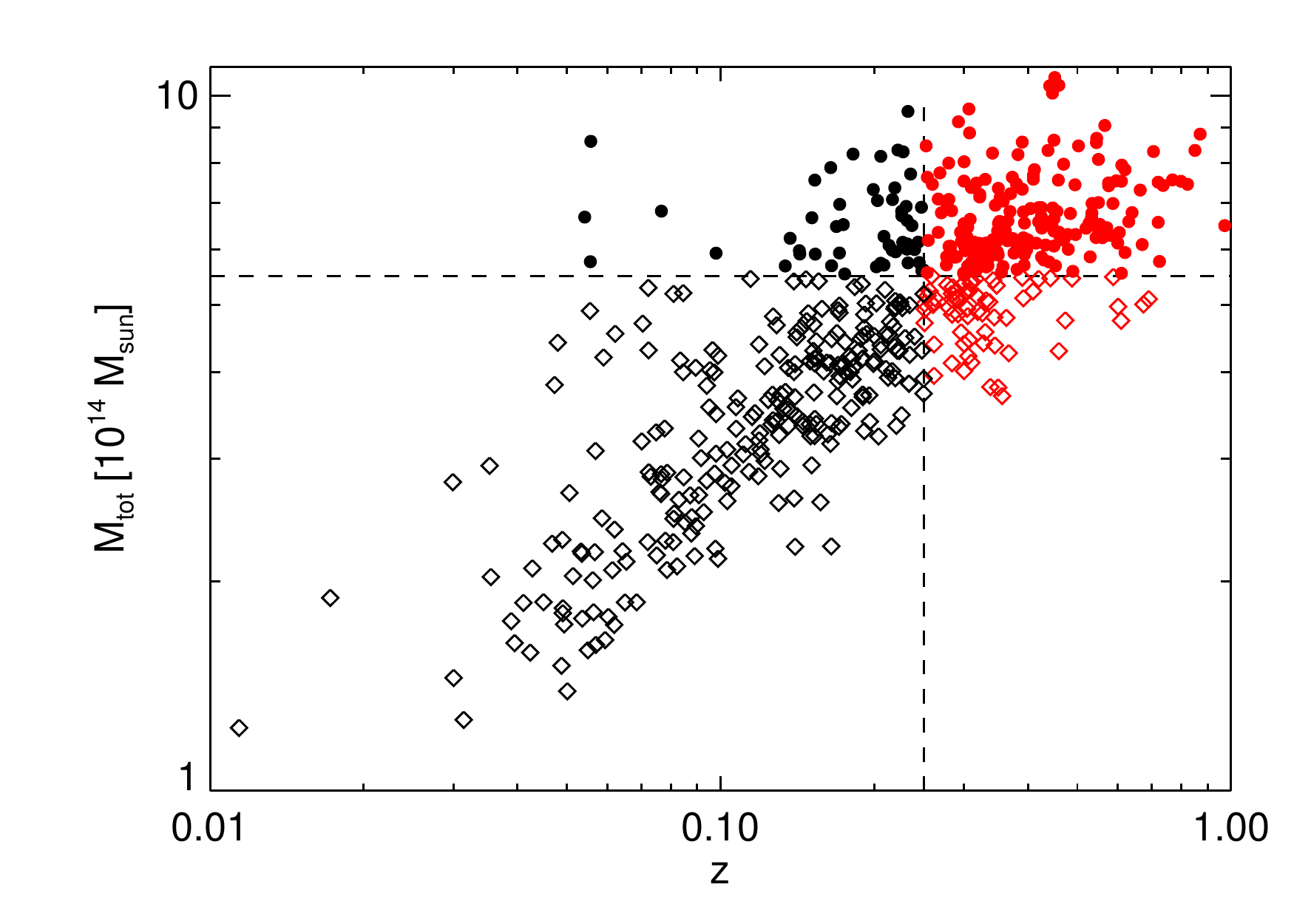}
    \caption{In the redshift--mass plane, we show the distribution of the 561
\Planck\ clusters with known redshift, which are used in this paper. Different
colours (black and red) are used for the low- and high-redshift sub-samples,
while different symbols (dots and diamonds) are used for the low- and high-mass
sub-samples.}
         \label{subsamples_M_z}
\end{figure}

For the low-redshift and low-mass sub-samples we have no detection at
$\nu\,{<}\,353\,$GHz. The SEDs for the two redshift sub-samples are consistent
with each other (within $\Delta F_{\rm b}$), even though fluxes are
systematically higher at higher $z$. This is expected since in more distant
(younger) clusters there will be more gas-rich active (i.e., star-forming)
spirals.  Conversely, nearby clusters mainly contain elliptical galaxies, with
a lower star-formation rate and little dust. The same trend in redshift (for
$z\,{<}\,0.15$ versus $z\,{>}\,0.15$) was also seen in \citet{planck2014-a29},
in which a similar stacking was performed on \Planck\ data to investigate the
cross-correlation between the tSZ effect and the
CIB fluctuations. Given that our integration radius is significantly larger
than the typical cluster size for both sub-samples ($\theta_{500}$,
Table~\ref{table:best_fit}), we can exclude the possibility that the larger
fluxes found for $z>0.25$ are biased high because of the smaller angular size
of more distant objects. For the high- and low-mass sub-samples, the difference
between the two SEDs becomes even more important, with higher fluxes when
$M^{500}_{\rm tot}>5.5\times10^{14}\,{\rm M}_{\odot}$. We expect dust emission
to be proportional to the total cluster mass, because they should both be
tightly correlated with the number of galaxies
\citep{GiardMontier2008, daSilva2009}.  A similar behaviour for the sub-samples
in $M^{500}_{\rm tot}$ and $z$ is not surprising. The two partitions do not
trace exactly the same populations, but they are not completely independent,
as shown in Fig.~\ref{subsamples_M_z}. The lower redshift bin strongly
overlaps with lower mass systems and vice versa.

\begin{table}[tbp!]
\begingroup
\newdimen\tblskip \tblskip=5pt
\caption{Fluxes found in the co-added maps for a sample of 645 clusters
extracted from the first Planck Catalogue of SZ sources. $\Delta F_{\rm b}$ is
the uncertainty estimated using bootstrap resampling, while $\Delta F$ is
obtained by integrating at random positions around the cluster and around the
regions centred 1\deg\ away from the cluster Galactic latitude and longitude
(see Sect.~\ref{tests} and Fig.~\ref{depointing}), and $\Delta F_{\rm ran}$
is determined the same way as $\Delta F_{\rm d}$ except replacing the cluster
position with random positions on the sky.}
\label{table:sed}
%\nointerlineskip
\vskip -1mm
\footnotesize
\setbox\tablebox=\vbox{
   \newdimen\digitwidth 
   \setbox0=\hbox{\rm 0} 
   \digitwidth=\wd0 
   \catcode`*=\active 
   \def*{\kern\digitwidth}
   \newdimen\dpwidth 
   \setbox0=\hbox{.} 
   \dpwidth=\wd0 
   \catcode`!=\active 
   \def!{\kern\dpwidth}
   \newdimen\signwidth 
   \setbox0=\hbox{+} 
   \signwidth=\wd0 
   \catcode`?=\active 
   \def?{\kern\signwidth}
\halign{\tabskip 0em\hbox to 2.0cm{#\leaderfil}\tabskip 1em&
     \hfil#\hfil \tabskip 1em&
     \hfil#\hfil \tabskip 1em&
     \hfil#\hfil \tabskip 1em&
     \hfil#\hfil \tabskip 0em\cr
\noalign{\doubleline}
\omit& $F$& $\Delta F_{\rm b}$& $\Delta F$& $\Delta F_{\rm ran}$\cr
\noalign{\vskip 2pt}
\omit& [Jy]& [Jy]& [Jy]& [Jy]\cr
\noalign{\vskip 3pt\hrule\vskip 5pt}
100\,GHz&  $-$0.0009& $\pm0.0031$& $\pm0.0028$& $\pm0.0031$\cr
143\,GHz&    ?0.0010& $\pm0.0030$& $\pm0.0023$& $\pm0.0034$\cr
217\,GHz&    ?0.0103& $\pm0.0092$& $\pm0.0056$& $\pm0.012*$\cr
353\,GHz&    ?0.098*& $\pm0.036*$& $\pm0.020*$& $\pm0.042*$\cr
545\,GHz&    ?0.34**& $\pm0.12**$& $\pm0.063*$& $\pm0.13**$\cr
853\,GHz&    ?0.94**& $\pm0.33**$& $\pm0.18**$& $\pm0.34**$\cr
100\,$\mu$m& ?0.86**& $\pm0.40**$& $\pm0.22**$& $\pm0.59**$\cr
60\,$\mu$m&  ?0.269*& $\pm0.097*$& $\pm0.083*$& $\pm0.34**$\cr
\noalign{\vskip 3pt\hrule\vskip 5pt}
}}
\endPlancktable 
\endgroup
\end{table}

The SEDs shown in Figs.~\ref{fig:sed} and \ref{subsamples} behave like Galactic
dust, confirming the hypothesis of thermal dust emission. They can be well
represented by modified blackbody emission (the black curve in
Fig.~\ref{fig:sed}), with spectral index $\beta$.
This accounts for the fact that the clusters are not perfect blackbodies, but
have a power-law dust emissivity
$\kappa_{\nu}=\kappa_{0}(\nu/\nu_0)^{\,\beta}$, i.e.,
\begin{equation}
 I_\nu = A_0 \left(\frac{\nu}{\nu_{0}}\right)^{\beta} B_\nu (T_{\rm d0}),
   \label{mbb_model}
\end{equation}
where $\beta$ is the emissivity index, $B_{\nu}$ is the Planck function,
$T_{\rm d0}$ the dust temperature, and $A_0$ an overall amplitude (directly
related to the dust mass, as will be discussed in Sect.~\ref{dust_mass}).
The combination of \Planck\ and IRAS spectral coverage allows us to sample the
SED of the average IR cluster across its emission peak. The reconstructed SED
can be used to constrain the dust temperature and also the amplitude $A_0$
(Eq.~\ref{mbb_model}). This model has the advantage of being accurate enough to
adequately fit the data, while providing a simple interpretation of the
observations, with a small number of parameters. Even though studies of
star-forming galaxies have demonstrated the inadequacy of a single temperature
model for detailed, high signal-to-noise SED shapes
\citep[e.g.,][]{DaleHelou,Wiklind2003},
the cold dust component ($\lambda\ga100\,\mu$m)
tends to be well represented using a single effective temperature modified
blackbody \citep{DraineLi2007,Casey2012,Clemens2013}. 
Fig.~\ref{fig:sed} shows that a further component would indeed permit us to
also fit the 60$\,\mu$m point (the rightmost point in Fig.~\ref{fig:sed}).
This excess at high frequency could be caused by smaller grains that are not
in thermal equilibrium with the radiation field; they are stochastically heated
and therefore their emission is not a simple modified blackbody
\citep{Compiegne2011, Jones2013}.  However, this additional contribution would
be sub-dominant at $\lambda\geq100\,\mu$m and so would not significantly change
the derived temperature and mass of the cold dust, which corresponds to the
bulk of the dust mass in galaxies \citep{Cortese2012,Davies2012,Santini2014}.
Even if the signal at
60$\,\mu$m indicates that a two-temperature dust model may eventually provide
a more accurate representation of the average cluster dust SED over the whole
frequency range explored here, we choose to adopt a single-component approach
and exclude the 60$\,\mu$m from the data used for the fit. Not doing this
would bias the estimate of the temperature and mass of the cold dust.
 
\subsection{Dust temperature}\label{temp}
By representing the recovered SEDs with a single-temperature modified
blackbody dust model and fixing $\beta$, we can constrain the SED and estimate
the average dust temperature.  This is for the observer's frame, while
the temperature in the rest-frame of the cluster will be given by
$T_{\rm d}=T_{\rm d0}(1+z)$.  We use a $\chi^2$ minimization approach and
account for colour corrections when
comparing the measured SED to the modelled one.

 \begin{table*}[tbp!]
\begingroup
\newdimen\tblskip \tblskip=5pt
\caption{Integrated fluxes from the stacked maps obtained, for each frequency,
by co-adding patches extracted at the positions of 307 clusters at
$z\le0.25$ and 254 clusters at $z>0.25$. As in Table~\ref{table:sed},
$\Delta F_{\rm b}$ is the uncertainty estimated using bootstrap resampling,
while $\Delta F_{\rm d}$ is obtained by integrating at random positions around
the cluster and around the regions centred 1\deg\ away from the cluster
Galactic latitude and longitude positions (see Sect.~\ref{tests} and
Fig.~\ref{depointing}).}
\label{table:sed_z}
\nointerlineskip
\vskip -1mm
\footnotesize
\setbox\tablebox=\vbox{
   \newdimen\digitwidth 
   \setbox0=\hbox{\rm 0} 
   \digitwidth=\wd0 
   \catcode`*=\active 
   \def*{\kern\digitwidth}
   \newdimen\dpwidth 
   \setbox0=\hbox{.} 
   \dpwidth=\wd0 
   \catcode`!=\active 
   \def!{\kern\dpwidth}
   \newdimen\signwidth
   \setbox0=\hbox{+}
   \signwidth=\wd0
   \catcode`?=\active
   \def?{\kern\signwidth}
\halign{\tabskip 0em\hbox to 2.5cm{#\leaderfil}\tabskip 2em&
     \hfil#\hfil\tabskip 2em&
     \hfil#\hfil\tabskip 2em&
     \hfil#\hfil\tabskip 2em&
     \hfil#\hfil\tabskip 2em&
     \hfil#\hfil\tabskip 2em&
     \hfil#\hfil\tabskip 0em\cr
\noalign{\doubleline}
\omit & \multispan3\hfil $z\le0.25$\hfil & \multispan3\hfil $z>0.25$\hfil\cr
\noalign{\vskip -2pt}
\omit & \multispan3\hrulefill& \multispan3\hrulefill\cr
\omit& $F$& $\Delta F_{\rm b}$& $\Delta F$& $F$& $\Delta F_{\rm b}$& $\Delta F$\cr
\omit& [Jy]& [Jy]& [Jy]& [Jy]& [Jy]& [Jy]\cr
\noalign{\vskip 3pt\hrule\vskip 5pt}
100\,GHz&  $-$0.0042& $\pm0.0047$& $\pm0.0034$& $-$0.0002& $\pm0.0046$& $\pm0.0034$\cr
143\,GHz&  $-$0.0032& $\pm0.0046$& $\pm0.0047$&   ?0.0037& $\pm0.0046$& $\pm0.0047$\cr
217\,GHz&    ?0.004*& $\pm0.015*$& $\pm0.0041$&   ?0.022*& $\pm0.016*$& $\pm0.0041$\cr
353\,GHz&    ?0.083*& $\pm0.056*$& $\pm0.012*$&   ?0.144*& $\pm0.070*$& $\pm0.012*$\cr
545\,GHz&    ?0.33**& $\pm0.18**$& $\pm0.049*$&   ?0.50**& $\pm0.22**$& $\pm0.049*$\cr
853\,GHz&    ?0.93**& $\pm0.53**$& $\pm0.14**$&   ?1.40**& $\pm0.62**$& $\pm0.14**$\cr
100$\,\mu$m& ?1.00**& $\pm0.60**$& $\pm0.38**$&   ?1.37**& $\pm0.69**$& $\pm0.38**$\cr
60$\,\mu$m&  ?0.26**& $\pm0.14**$& $\pm0.46*$&    ?0.44**& $\pm0.17**$& $\pm0.46**$\cr
\noalign{\vskip 3pt\hrule\vskip 5pt}
}}
\endPlancktable
\endgroup
\end{table*}

\begin{table*}[tbp!]
\begingroup
\newdimen\tblskip \tblskip=5pt
\caption{Integrated fluxes from the stacked maps obtained, for each frequency,
by co-adding patches extracted at the positions of 320 clusters with
$M^{500}_{\rm tot}\leq5.5\times10^{14}\,{\rm M}_{\odot}$ and 241 clusters with
$M^{500}_{\rm tot}>5.5\times10^{14}\,{\rm M}_{\odot}$. As in
Table~\ref{table:sed}, $\Delta F_{\rm b}$ is the uncertainty estimated with
bootstrap resampling, while $\Delta F$ is obtained by integrating at random
positions around the cluster and around regions centred 1\deg\ away from the
cluster Galactic latitude and longitude positions (see Sect.~\ref{tests} and
Fig.~\ref{depointing}).}
\label{table:sed_m}
\nointerlineskip
\vskip -1mm
\footnotesize
\setbox\tablebox=\vbox{
   \newdimen\digitwidth 
   \setbox0=\hbox{\rm 0} 
   \digitwidth=\wd0 
   \catcode`*=\active 
   \def*{\kern\digitwidth}
   \newdimen\dpwidth 
   \setbox0=\hbox{.} 
   \dpwidth=\wd0 
   \catcode`!=\active 
   \def!{\kern\dpwidth}
   \newdimen\signwidth
   \setbox0=\hbox{+}
   \signwidth=\wd0
   \catcode`?=\active
   \def?{\kern\signwidth}
\halign{\tabskip 0em\hbox to 2.5cm{#\leaderfil}\tabskip 2em&
     \hfil#\hfil\tabskip 2em&
     \hfil#\hfil\tabskip 2em&
     \hfil#\hfil\tabskip 2em&
     \hfil#\hfil\tabskip 2em&
     \hfil#\hfil\tabskip 2em&
     \hfil#\hfil\tabskip 0em\cr
\noalign{\doubleline}
\omit& \multispan3\hfil $M^{500}_{\rm tot}\le5.5\times10^{14}\,{\rm M}_{\odot}$\hfil& \multispan3\hfil $M^{500}_{\rm tot}>5.5\times10^{14}\,{\rm M}_{\odot}$\hfil\cr
\noalign{\vskip -5pt}
\omit & \multispan3\hrulefill& \multispan3\hrulefill\cr
\omit& $F$& $\Delta F_{\rm b}$& $\Delta F$& $F$& $\Delta F_{\rm b}$& $\Delta F$\cr
\omit& [Jy]& [Jy]& [Jy]& [Jy]& [Jy]& [Jy]\cr
\noalign{\vskip 3pt\hrule\vskip 5pt}
100\,GHz&  $-$0.0036& $\pm0.0047$& $\pm0.0034$& $-$0.0015& $\pm0.0054$& $\pm0.0034$\cr
143\,GHz&  $-$0.0031& $\pm0.0042$& $\pm0.0047$&   ?0.0034& $\pm0.0055$& $\pm0.0047$\cr
217\,GHz&  $-$0.0033& $\pm0.0126$& $\pm0.0041$&   ?0.031*& $\pm0.018*$& $\pm0.0041$\cr
353\,GHz&    ?0.035*& $\pm0.048*$& $\pm0.012*$&   ?0.214*& $\pm0.079*$& $\pm0.012*$\cr
545\,GHz&    ?0.152*& $\pm0.155*$& $\pm0.049*$&   ?0.75**& $\pm0.25**$& $\pm0.049*$\cr
853\,GHz&    ?0.45**& $\pm0.44**$& $\pm0.14**$&   ?2.10**& $\pm0.71**$& $\pm0.14**$\cr
100\,$\mu$m& ?0.60**& $\pm0.52**$& $\pm0.38**$&   ?1.93**& $\pm0.80**$& $\pm0.38**$\cr
60\,$\mu$m&  ?0.16**& $\pm0.12**$& $\pm0.46**$&   ?0.57**& $\pm0.19**$& $\pm0.46**$\cr
\noalign{\vskip 3pt\hrule\vskip 5pt}
}}
\endPlancktable 
\endgroup
\end{table*}

\begin{table*}[tbp!]
\begingroup
\newdimen\tblskip \tblskip=5pt
\caption{For the different cluster samples considered here, we provide the
average redshifts ($z$), characteristic radii ($\theta_{500}$) and total masses
at a radius for which the mean cluster density is 500 and 200 times the
critical density of the Universe ($M^{500}_{\rm tot}$ and $M^{200}_{\rm tot}$).
The best-fit temperature and dust mass are also provided, exploring different
choices for the emissivity index $\beta$ for the full sample of 645 clusters.}
\label{table:best_fit}
\nointerlineskip
\vskip -1mm
\footnotesize
\setbox\tablebox=\vbox{
   \newdimen\digitwidth 
   \setbox0=\hbox{\rm 0} 
   \digitwidth=\wd0 
   \catcode`*=\active 
   \def*{\kern\digitwidth}
   \newdimen\dpwidth 
   \setbox0=\hbox{.} 
   \dpwidth=\wd0 
   \catcode`!=\active 
   \def!{\kern\dpwidth}
   \newdimen\signwidth
   \setbox0=\hbox{+}
   \signwidth=\wd0
   \catcode`?=\active
   \def?{\kern\signwidth}
\halign{\tabskip 0em\hbox to 5.0cm{#\leaderfil}\tabskip 2em&
     \hfil#\hfil\tabskip 1em&
     \hfil#\hfil\tabskip 1em&
     \hfil#\hfil\tabskip 1em&
     \hfil#\hfil\tabskip 1em&
     \hfil#\hfil\tabskip 1em&
     \hfil#\hfil\tabskip 1em&   
     \hfil#\hfil\tabskip 0em\cr
\noalign{\doubleline}
\omit\hfil Sample\hfil& $\langle z\rangle$& $\langle\theta_{500}\rangle$& $\langle M^{500}_{\rm tot}\rangle$& $\langle M^{200}_{\rm tot}\rangle$& $\beta$& $T_{\rm d0}$ [K]& $M_{\rm dust}$\cr
\noalign{\vskip 2pt}
\omit& \omit& [arcmin]& [$10^{14}\,{\rm M}_{\odot}$]& [$10^{14}\,{\rm M}_{\odot}$]& & [K]& [$10^{10}\,{\rm M}_{\odot}$]\cr 
\noalign{\vskip 3pt\hrule\vskip 5pt}
Full sample& $\mathbf{0.26*\pm0.17*}$& $\mathbf{7.4*\pm5.3}$& $\mathbf{5.1\pm1.9}$& $\mathbf{5.6\pm2.1}$& $\mathbf{1.5}$& $\mathbf{19.2\pm2.4}$& $\mathbf{1.08\pm0.32}$\cr
\noalign{\vskip 2pt}
\omit& & & & & 2.2& $15.3\pm1.2$& $1.25\pm0.36$\cr 
\omit& & & & & 2.1& $15.7\pm1.2$& $1.24\pm0.34$\cr 
\omit& & & & & 2.0& $16.2\pm1.4$& $1.25\pm0.37$\cr
\omit& & & & & 1.9& $16.7\pm1.5$& $1.24\pm0.37$\cr
\omit& & & & & 1.8& $17.2\pm1.6$& $1.20\pm0.35$\cr
\omit& & & & & 1.7& $17.8\pm1.8$& $1.17\pm0.36$\cr
\omit& & & & & 1.6& $18.5\pm2.1$& $1.14\pm0.34$\cr
\omit& & & & & 1.4& $20.0\pm2.8$& $1.03\pm0.32$\cr
\omit& & & & & 1.3& $20.9\pm3.3$& $0.97\pm0.32$\cr
\noalign{\vskip 3pt\hrule\vskip 5pt}
$z\le0.25$ (307)& $0.139\pm0.063$& $9.6*\pm6.3*$& $4.0\pm1.6$& $4.3\pm1.7$& 1.5& $20.7\pm9.9$& $0.34\pm0.17$\cr 
$z>0.25$ (254)& $0.41*\pm0.13*$& $4.68\pm0.38$& $6.4\pm1.3$& $7.0\pm1.5$& 1.5& $19.2\pm3.2$& $2.56\pm0.91$\cr
\noalign{\vskip 3pt\hrule\vskip 5pt}
$M^{500}_{\rm tot}\le5.5\times10^{14}\,{\rm M}_{\odot}$ (320)& $0.17*\pm0.11*$& $9.1*\pm6.4*$& $3.7\pm1.1$& $4.1\pm1.2$& 1.5& $20.3\pm5.1$& $0.21\pm0.14$\cr
\noalign{\vskip 2pt}
$M^{500}_{\rm tot}>5.5\times10^{14}\,{\rm M}_{\odot}$ (241)& $0.38*\pm0.16*$& $5.2*\pm1.7*$& $6.8\pm1.1$& $7.5\pm1.2$& 1.5& $19.2\pm2.0$& $3.48\pm0.99$\cr
\noalign{\vskip 3pt\hrule\vskip 5pt}
}}
\endPlancktable 
\endgroup
\end{table*}

In \cite{planck2014-XXII} the \Planck-HFI intensity (and polarization) maps
were used to estimate the spectral index $\beta$ of the Galactic dust emission.
On the basis of nominal mission data, they found that, at $\nu<353\,$GHz, the
dust emission can be well represented by a modified blackbody spectrum with
$\beta=1.51\pm0.06$. At higher frequencies ($100\,\mu$m--353\,GHz)
$\beta=1.65$ was assumed. In the following we have adopted a single spectral
index over the whole spectral range, and $\beta=1.5$ will be our baseline
value. In some other studies an emissivity index $\beta=2$ has been used
instead, for example in the analysis of \cite{Davies2012}, which focused on
\textit{Herschel} data (at 100--500$\,\mu$m) to explore the IR properties of
cluster galaxies (specifically 78 galaxies in the Virgo cluster). The spectral
index $\beta$ is known to vary with environment. Shown by \Planck\ to be
equal to 1.50 for the diffuse ISM in the Solar neighbourhood, $\beta$ is known
to be higher in molecular clouds \citep{planck2013-p06b}. This reflects
variations in the composition and structure of dust, something that is clearly
seen in laboratory measurements \citep[e.g.,][]{Jones2013}.

The impact of the
adopted emissivity index on our results is explored by varying its value
between 1.3 and 2.2. For the total sample of 645 clusters, we have explored how
this affects $T_{\rm d0}$ and $M_{\rm dust}$, and we summarize the results in
Table~\ref{table:best_fit}. When we
compare the dust temperature obtained with the same choice of $\beta$,
i.e., $\beta=2$, the dust temperatures that we find are similar to those obtained by 
\cite{Davies2012} for the galaxies of the Virgo cluster. The range explored shows that when reducing
$\beta$ the inferred $T_0$ increases, and vice versa. However, the differences
are not significant within the associated uncertainties, reflecting the fact
that our data are not good enough to simultaneously constrain the temperature,
the amplitude, and also the spectral index. Different values of $\beta$ can
affect the dust mass estimates by up to 20\,\%, which is negligible with
respect to the existing uncertainty on the dust opacity $\kappa_{\nu}$ (as will
be discussed in Sect.~\ref{dust_mass}).
 
In Table~\ref{table:best_fit} we report the best-fit values that we obtain for
$T_{\rm d0}$ and $M_{\rm dust}$\footnote{We report $M_{\rm dust}$ rather than
$A_0$, the overall amplitude to which it is proportional, see
Sect.~\ref{dust_mass}.} when considering the sample of 645 clusters, as well as
the two redshift and mass sub-samples. The associated uncertainties are derived
from the statistical ones obtained through the $\chi^2$ minimization in the
fitting procedure (with $\Delta F_{\rm b}$ being the error on the flux at each
frequency). The corresponding best-fit models ($\beta=1.5$) are represented in
Figs.~\ref{fig:sed} (by the black solid line) and \ref{subsamples} (blue dashed
and orange dot-dashed lines). 

Figure~\ref{fig:sed} shows that the 143 and 353\,GHz intensities are slightly
lower and higher, respectively, with respect to the best-fit model. This might
indicate a residual tSZ contamination, since the tSZ signal is negative at
143\,GHz and positive at 353\,GHz. The SZ amplitude that we subtract has been
estimated under the non-relativistic hypothesis, and this could result in a
slight underestimation of the cluster Comptonization parameter $y$.  Although the 
relativistic correction is expected to have a small impact on the inferred Compton parameter, 
we should note that for cluster temperatures of a few Kelvin the relativistic correction 
boosts the tSZ flux at 857 by a factor of several tens of a percent. Given the amplitude 
of the SZ contribution at this frequency with respect to the dust emission, this contribution 
remains negligible. On the other hand, the cluster IR component we are considering here is 
not included in none of the simulations used to test the 
{\tt MILCA} algorithm and might bias somehow the reconstruction of the tSZ amplitude.
Then we have verified that the residual tSZ contamination is negligible by fitting, a
posteriori, the dust SED as a linear combination of the modified blackbody
and a tSZ contribution. For the two components we find an amplitude consistent
with 1 and 0, respectively.
For the whole sample of 645 objects, we assume that the mean redshift is the
mean of the known redshifts ($z=0.26\pm0.17$). We then obtain
$T_{\rm d}=(24.2\pm3.0\pm2.8)\,$K, where the additional systematic uncertainty
is due to the redshift dispersion of our sample (see the second column of
Table~\ref{table:best_fit}).  We observe a slight increase of dust temperature
with redshift, obtaining $T_{\rm d}(z\,{\le}\,0.25)=(24\pm11)\,$K and
$T_{\rm d}(z\,{>}\,0.25)=(27.1\pm4.5)\,$K, but no significant evolution within
the uncertainties.
For the low- and high-mass sub-samples we have
$T_{\rm d}(M^{500}_{\rm tot}\,{\leq}\,5.5\times10^{14}{\rm M}_{\odot})
=(23.7\pm6)\,$K and
$T_{\rm d}(M^{500}_{\rm tot}\,{>}\,5.5\times10^{14}{\rm M}_{\odot})
=(26.4\pm4.5)\,$K, respectively. The recovered dust temperatures are in
agreement with those observed for the dust content in various field galaxy
samples \citep[e.g.,][]{Dunne2011,Clemens2013,Symeonidis2013} and with the
values expected for the cold dust component in cluster galaxies, e.g., the
Virgo sample explored by \cite{Davies2012, SeregoAlighieri2013}.

\subsection{Dust mass}\label{dust_mass}
Following the prescription of \cite{Hildebrand1983}, the dust mass can be
estimated from the observed flux densities and the modified blackbody
temperature as
\begin{equation}
  M_{\rm d}= \frac{S_{\nu}D^2(1+z) K}{\kappa_{\nu} B_{\nu}(T_{\mathrm{d0}})},
\end{equation}
with the ``K-correction'' being
\begin{equation}
	K=\left(\frac{\nu_{\rm obs}}{\nu_{\rm em}}\right)^{3+\beta}
 \frac{e^{(h\nu_{\rm em}/kT_{\rm d0})}-1}{e^{(\nu_{\rm obs}/kT_{\rm d0})}-1},
\end{equation}
which allows translation to monochromatic rest-frame flux densities, $S_{\nu}$.
Here $\nu_{\rm obs}$ represents the observed frequency and $\nu_{\rm em}$ the
rest-frame frequency, with $\nu_{\rm obs}=\nu_{\rm em}/(1+z)$, and
$D$ is the radial comoving distance. The amplitude parameter $A_{\rm 0}$ (see
Eq.~\ref{mbb_model}) then provides an estimate of the overall dust mass:
\begin{equation}
	M_{\rm d} =  \frac{A_0}{\kappa_{\nu_0}} \Omega D^2(1+z) K,
	\label{m_dust}
\end{equation}
provided we know $\kappa_{\nu_0}$, the dust opacity at a given frequency
$\nu_0$, with $\Omega$ being the solid angle. Here we adopt the value
$\kappa_{850}=0.0383\,{\rm m}^2\,{\rm kg}^{-1}$ \citep{Draine2003}.

Dust mass estimates obtained from Eq.~(\ref{m_dust}) are also listed in
Table~\ref{table:best_fit}. The corresponding uncertainties have been derived
using random realizations of the model, letting $T_{\rm d0}$ and $A_0$ vary
within the associated uncertainties and accounting for the correlation
between the two.

For the whole sample we obtain an average dust mass of
$(1.08\pm0.32)\times10^{10}\,{\rm M}_{\odot}$. Our estimates are similar
to those obtained with different approaches by \cite{Muller2008}
($M_{\rm d}=8\times10^9\,{\rm M}_{\odot}$) for a sample with a comparable
redshift distribution and \cite{Gutierrez2014}
($M_{\rm d}<8.4\times10^9\,{\rm M}_{\odot}$) for a relatively low-mass cluster
sample. For the low- and high-mass sub-samples we find
$(0.21\pm0.14)\times10^{10}\,{\rm M}_{\odot}$ and
$(3.48\pm0.99)\times10^{10}\,{\rm M}_{\odot}$, respectively.
This implies that the dust mass is responsible for the difference between the
two curves in the right panel of Fig.~\ref{subsamples}. The similar trend in
mass and redshift, i.e., $(0.34\pm0.17)\times10^{10}\,{\rm M}_{\odot}$, when
$z\le0.25$ versus $(2.56\pm0.91)\times10^{10}\,{\rm M}_{\odot}$, when $z>0.25$,
is due to the fact that most of the low-mass objects are detected at low
redshift and vice versa.  

The \cite{Draine2003} value of $\kappa_{850}$ was derived from a dust model
(including assumptions about chemical composition, distribution of grain size,
etc.) that shows good agreement with available data.  However, using a
different approach, \cite{James2002} found
$\kappa_{850}=(0.07\pm0.02)\,{\rm m}^2\,{\rm kg}^{-1}$, nearly a factor of 2
higher. The latter value uses a calibration that has been obtained with a
sample of galaxies for which the IR fluxes, gas masses, and metallicities
are all available, and adopts the assumption that the fraction of metals
bound up in dust is constant for all the galaxies in the sample. From just
these two published results, it appears that the normalization of the dust
opacity represents the major source of uncertainty in deriving dust masses
from observed IR fluxes \citep{Fanciullo2015}. In fact, \cite{planck2014-XXIX}
has shown that the opacity of the \cite{Draine2003} model needs to be increased
by a factor of 1.8 to fit the \Planck\ data, as well as extinction
measurements, and this moves the normalization towards the value suggested
by \cite{James2002}.

\subsection{Dust-to-gas mass ratio}
We now estimate the ratio of dust mass to gas mass, $Z_{\rm d}$, directly from
the observed IR cluster signal.  To do so we need an estimate of the cluster
gas mass.  The IR fluxes used in the previous section to estimate $M_{\rm d}$
were obtained by integrating the signal out to 15$\arcm$ from the cluster
centres, without any rescaling of the maps with respect to each cluster's
characteristic radius. In the PSZ1, the mass information is provided at
$\theta_{500}$, and for our sample this size is significantly smaller than
15$\arcm$ for almost all clusters (see Table~\ref{table:best_fit}). However,
we can take advantage of the self-similarity of cluster profiles, from which it
follows that the ratio between radii corresponding to different overdensities
is more or less constant over the cluster population. Therefore we will use
$\theta_{200}$ and assume that $\theta_{200}\simeq1.4\times\theta_{500}$
\citep{Ettori2009} to obtain the corresponding enclosed total mass
$M^{200}_{\rm tot}=(4\pi \rho_{\rm c} \theta_{200}^3)/3$. Here, $\theta_{500}$
is obtained from the $M^{500}_{\rm tot}$ values listed in the PSZ1. The cluster
total mass also provides an estimate of the gas mass,
$M^{200}_{\rm gas}\simeq0.1\,M^{200}_{\rm tot}$
\citep[e.g.,][]{Pratt2009, Comis2011, planck2012-V, Sembolini2013}.
Assuming that $M_{\rm d}$ and $M^{200}_{\rm tot}$ correspond to comparable
cluster regions, we find that $Z_{\rm d}=(1.93\pm0.92)\times10^{-4}$ for the
full sample. For the low- and high-redshift sub-samples we find
$(0.79\pm0.50)\times10^{-4}$ and $(3.7\pm1.5)\times10^{-4}$, respectively,
while for the low- and high-mass sub-samples we obtain
$(0.51\pm0.37)\times10^{-4}$ and $(4.6\pm1.5)\times10^{-4}$, respectively,
Note that the uncertainties quoted here do not account for the fact that the
gas fraction might vary from cluster to cluster, and as a function of
radius/mass and redshift. 

These dust-to-mass ratios are derived from the overall IR flux, which is the
sum of the contribution due to the cluster galaxies and a possible further
contribution coming from the ICM dust component. Therefore they represent an
upper limit for the dust fraction that is contained within the IGM/ICM. These
values are consistent with the upper limit ($5\times10^{-4}$) derived by
\citet{GiardMontier2008} from the $L_{\rm IR}/L_{\rm X}$ ratio and a model for
the ICM dust emission \citep{Montier2004}. 

%% file: 05_conclusions.tex
We have adopted a stacking approach in order to recover the average SED
of the IR emission towards galaxy clusters. 
Considering the Catalogue of Planck SZ Clusters, we have used the \Planck-HFI
maps (from 100 to 857\,GHz) as well as the IRAS maps (IRIS data at 100 and
60$\,\mu$m) in order to sample the SED of the cluster dust emission on both
sides of the expected emissivity peak. 

For a sample of 645 clusters selected from the PSZ1 catalogue, we find
significant detection of dust emission from 353 to 857\,GHz, as well as
at 100$\,\mu$m and 60$\,\mu$m, at the cluster positions, after cleaning for
foreground contributions. By co-adding maps extracted at random positions on
the sky, we have verified that the residual Galactic emission is accounted for
in the uncertainty budget. For the IRAS frequencies, we find average central
intensities that are in agreement with what was found by
\citet{MontierGiard2005}. The 143\,GHz and 353\,GHz data may be slightly
affected by residual tSZ contamination. Although the dust component contributes 
to the high frequency bands in \Planck, its impact on the reconstruction of the tSZ 
amplitude is expected to be very small. However we have verified a posteriori that 
this does not impact our results significantly. The measured SED is consistent with 
dust IR emission following a modified blackbody distribution.

These results have allowed us
to constrain, directly form its IR emission, both the average overall dust
temperature and the dust mass in clusters. From the average cluster SED we
infer an average dust temperature of $T_{\rm d}=(24.2\pm3.0)\,$K, in agreement
with what is observed for Galactic thermal dust emission. 
 The average dust temperature, as estimated for the total
sample of 645 clusters, leads to an average dust mass of
$M_{\rm d}=(1.25\pm0.37)\times10^{10}\,{\rm M}_{\odot}$. 
By dividing our initial sample of 645 objects into two bins, according to
either their mass or redshift, we find that the IR emission is larger for the
higher mass (or higher redshift) clusters. This difference is mainly due to
a larger dust mass ($(0.21\pm0.14)\times10^{10}\,{\rm M}_{\odot}$ versus
$(3.48\pm0.99)\times10^{10}\,{\rm M}_{\odot}$), the recovered temperatures
being instead consistent with each other. However, we stress that our sample
is not ideal for constraining the mass and redshift evolution of the IR
emission of the cluster dust component, since it is not complete and not
characterized by a well characterized selection function. Furthermore, the
redshift and mass bins, although not tracing exactly the same cluster
population, are not independent either.  With a larger sample, and a wider
distribution in mass and redshift, the separate mass and redshift dependance
could studied much more thoroughly, perhaps by correlating the weak signals
from individual clusters with $M_{\rm tot}$ and $z$. This approach would
allows us to better account for different distances, masses, and selection
effects.

Using the total mass estimates for each cluster, we derive the average cluster
total mass and so the dust-to-gas mass ratio
$Z_{\rm d}=(1.93\pm0.92)\times10^{-4}$. This leads to un upper limit on the
dust fraction within the ICM that is consistent with previous results.
Most of the IR signal detected in the maps stacked at cluster positions was expected to be due
to the contribution of the member galaxies \citep[e.g.][]{Roncarelli2010}. And the recovered temperature, typical
of values found in the discs of galaxies, is in agreement with this. However,
if we also take into account the additional uncertainties on the dust mass
estimates coming from the spectral index and the dust opacity
(up to 20\,\% and 50\,\%, respectively), our results cannot exclude a dust fraction that, according to
\citealt{Montier2004}), would imply that the IR ICM dust emission is an
important factor in the cooling of the cluster gas. 